\documentclass[submission,Phys]{SciPost}

\usepackage[english]{babel}
\usepackage[T1]{fontenc}
\usepackage{hyperref}

\usepackage{physics}
\usepackage{mathtools}
\usepackage{latexsym}
\usepackage{graphicx}
\usepackage{float}
\usepackage{dcolumn}
\usepackage{bm}
\usepackage{amsmath}
\usepackage{mathtools}
\usepackage[space]{grffile}
\usepackage{xcolor}
\usepackage{comment}
\usepackage{overpic}
\usepackage{subcaption}
\usepackage{ulem}

\def\XXint#1#2#3{{\setbox0=\hbox{$#1{#2#3}{\int}$}
     \vcenter{\hbox{$#2#3$}}\kern-.5\wd0}}

\newcommand{\gpol}{{\gamma_{\rm pol}}}

\def\XXint#1#2#3{{\setbox0=\hbox{$#1{#2#3}{\int}$}
     \vcenter{\hbox{$#2#3$}}\kern-.5\wd0}}

\hypersetup{hidelinks}

\usepackage[bitstream-charter]{mathdesign}
\fancypagestyle{SPstyle}{
\fancyhf{}
\lhead{\colorbox{scipostblue}{\bf \color{white} ~SciPost Physics }}
\rhead{{\bf \color{scipostdeepblue} ~Submission }}

\fancyfoot[C]{\textbf{\thepage}}
}

\begin{document}

\begin{center}{\Large \textbf{\color{scipostdeepblue}{
%%%%%%%%%% TODO: Write your article's title here
Emptiness formation in the Lieb--Liniger gas:\\
hydrodynamic instantons and a conjectured rate function
}}}\end{center}

\begin{center}
Boris A. Khanikati\textsuperscript{$\star$} and Alexander G. Abanov
\end{center}

\begin{center}
Department of Physics and Astronomy, Stony Brook University,\\
Stony Brook, New York 11794, USA\\
${}^\star$\small\sf bkhanikaev@g.harvard.edu
\end{center}

\section*{\color{scipostdeepblue}{Abstract}}
{\boldmath\textbf{%
We study the emptiness formation probability (EFP) in the ground state of the repulsive one-dimensional Lieb--Liniger Bose gas. For a macroscopic empty interval of length \(2R\), its leading asymptotic behavior is described by a rate function \(f(\gamma_0)\), defined by \(-\log P(R)\sim(\rho_0R)^2f(\gamma_0)\), where \(\rho_0\) is the mean density and \(\gamma_0\) is the dimensionless interaction strength. We propose a parameter-free integral equation for \(f(\gamma_0)\). Starting from the exact dual-field Fredholm-determinant representation of the EFP, we show how the conjectured kernel arises formally and identify the uniform asymptotic statement that remains to be proven for a rigorous derivation. The conjecture reproduces the Tonks--Girardeau and weak-coupling limits, as well as the first correction obtained independently in both limits. It also agrees at the few-percent level with numerical minimization of the Lieb--Liniger hydrodynamic action over more than four orders of magnitude in coupling. The numerical calculation yields the corresponding emptiness instantons and their astroid-like vacuum regions.
}}

%%%%%%%%%% BLOCK: Copyright information
% This block will be filled during the proof stage, and finilized just before publication.
% It exists here only as a placeholder, and should not be modified by authors.
%\noindent\textcolor{white!90!black}{%
%\fbox{\parbox{0.975\linewidth}{%
%\textcolor{white!40!black}{\begin{tabular}{lr}%
%  \begin{minipage}{0.6\textwidth}%
%    {\small Copyright attribution to authors. \newline
%    This work is a submission to SciPost Physics. \newline
%    License information to appear upon publication. \newline
%    Publication information to appear upon publication.}
%  \end{minipage} & \begin{minipage}{0.4\textwidth}
%    {\small Received Date \newline Accepted Date %\newline Published Date}%
%  \end{minipage}
%\end{tabular}}
%}}
%}
%%%%%%%%%% BLOCK: Copyright information

% For convenience during refereeing we turn on line numbers:
%\linenumbers

\vspace{10pt}
\noindent\rule{\textwidth}{1pt}
\tableofcontents
\noindent\rule{\textwidth}{1pt}
\vspace{10pt}

%%%%%%%%%%%%%%%%%%%%%%%
\section{Introduction}
\label{sec:Intro}
%%%%%%%%%%%%%%%%%%%%%%%
In recent years, rare fluctuations in many-body systems have attracted significant attention \cite{arzamasovs2019full, yeh_emptiness_2020, yeh2022emptiness, abanov2025polytropicEFP}, in part due to advances in experimental techniques for ultracold atomic gases \cite{bakr_quantum_2009, sherson_single-atom-resolved_2010, haller_single-atom_2015, parsons_site-resolved_2016, wei_quantum_2022}. A key quantity of interest is the \textit{emptiness formation probability} (EFP), particularly for integrable systems. First introduced in integrable spin chains as a correlator admitting a determinant representation \cite{korepin1994efp}, the EFP is the probability of finding no particles inside a spatial region of radius $R$ in the ground state of a many-body system. In several integrable models, the EFP can be studied analytically using the Bethe ansatz and related integrable methods such as exact multiple-integral representations and asymptotic limits for large emptiness \cite{kitanine2002xxz,kitanine2002XXZ-asymptotic,korepin2003xxz,colomo2008emptiness}. 

Among integrable models, the one-dimensional (1D) Bose gas with delta interactions, or the Lieb--Liniger model, serves as a paradigmatic example, particularly because of its experimental realizability \cite{Paredes2004, Kinoshita2004}, including measurements of correlation functions \cite{Tolra2004, Haller2011, Kinoshita2005}. As such, the model provides a natural setting to study the EFP. We consider here the repulsive model, 
\begin{equation}
    H = - \frac{\hbar^2}{2m} \sum_{i=1}^N \frac{\partial^2}{\partial {x_i}^2} + \frac{\hbar^2 }{2m} (2c)\sum_{i<j}\delta(x_i - x_j), \qquad c>0. 
\end{equation}
Henceforth, we work in natural units $\hbar = m = 1$. It is useful to characterize the interaction strength by the dimensionless coupling $\gamma = c/\rho$. At large emptiness size $R$, the leading term of the EFP can be expressed as a function of the dimensionless coupling at mean density $\gamma_0 = c / \rho_0$, 
\begin{equation}
    - \log P(R; \gamma_0) \sim (\rho_0 R)^2 f(\gamma_0) + o(R^2).
    \label{eq:rate-function}
\end{equation}

Despite substantial progress in determining the EFP for free fermions, spin chains, and several classes of quantum fluids \cite{abanov2002probability,AbanovFranchini2003,
FranchiniAbanov2005,abanov-hydro,yeh2022emptiness,
abanov2025polytropicEFP}, a general analytical expression for the Lieb--Liniger model at arbitrary coupling has remained a challenging problem. To the best of our knowledge, the only previous proposal for the leading large-emptiness asymptotics at arbitrary repulsive coupling is the conjecture of Korepin, Its, and Waldron (KIW) \cite{korepin1995probabilityphaseseparationbose}. Their work also established an exact dual-field Fredholm-determinant representation of the EFP, providing a natural starting point for asymptotic analysis by integrable-operator and Riemann–Hilbert methods. 

An alternative approach, introduced in Ref.~\cite{abanov-hydro}, determines the EFP from a saddle-point solution of the hydrodynamic equations in imaginary time subject to the emptiness boundary conditions. The action evaluated on this saddle gives the leading EFP exponent, while the saddle itself describes the imaginary-time evolution and spacetime shape of the empty region.
Despite the generality of this approach, finding exact analytical solutions remains challenging. This difficulty is partly tied to the implicit structure of the thermodynamic Bethe ansatz: while the equation of state is fully determined, it is given in terms of an integral equation that does not admit a general closed-form solution \cite{lieb1963exact, korepin1997quantum}.

At the same time, hydrodynamics has proven to be a successful tool for studying the EFP. Numerical solutions of the hydrodynamic equations revealed a parametric disagreement with the KIW conjecture in the weak-coupling limit, $c\to0$ \cite{yeh_emptiness_2020}. More recently, the emptiness instanton was solved analytically for the general class of polytropic fluids with an equation of state $\epsilon(\rho)\propto\rho^{\gpol}$. This result is particularly relevant to the Lieb--Liniger gas, whose equation of state approaches a polytropic form with $\gpol=2$ and $\gpol=3$ in the weak- and strong-coupling limits, respectively. The resulting weak-coupling asymptotic independently confirms the parametric disagreement with the KIW conjecture.

In this work, we revisit the KIW dual-field representation and propose a modified conjecture for the leading EFP exponent at arbitrary repulsive coupling. Starting from their exact dual-field determinant representation, we express the vacuum expectation as a Gaussian functional and organize the large-distance limit as a joint saddle of the Fredholm determinant and the dual field. We show that terms subleading in the determinant asymptotics for a fixed field can contribute at order \(R^2\) after the vacuum average because the relevant dual fields scale as \(O(R)\). Our proposal treats this conjectural asymptotic step as a uniform double-scaling problem.

The resulting conjecture expresses the leading EFP exponent in terms of the solution of an integral equation. It reproduces the known weak- and strong-coupling limits, including the first nontrivial correction in each regime, obtained independently through hydrodynamics. We further test the conjecture by numerically minimizing the hydrodynamic action for the Lieb--Liniger equation of state over more than four orders of magnitude in coupling. The numerical rate function closely follows the conjecture throughout the interaction crossover, while the corresponding minimizers provide the emptiness instantons and their astroid-like vacuum regions.

%%%%%%%%%%%%%%%%
\subsection*{Summary of main results}
\begin{figure}[t]
    \centering    \includegraphics[width=0.8\linewidth]{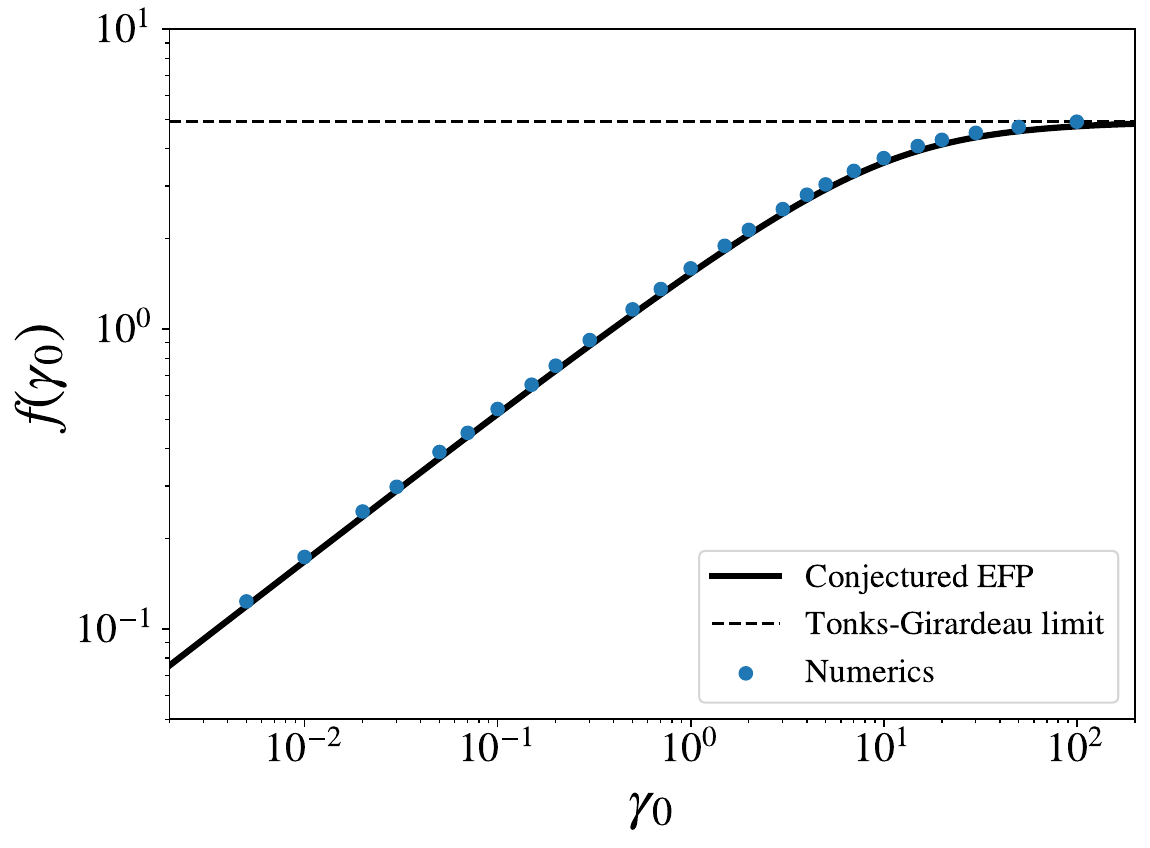}
    \caption{Comparison of the conjectured EFP rate function \(f(\gamma_0)\), obtained from Eqs.~\eqref{eq:P(r)} and \eqref{eq:dimless-int-eq}, with the numerical minimization of the hydrodynamic action. The solid line represents the conjecture, while the points show the numerical results. Both \(f(\gamma_0)\) and the dimensionless Lieb--Liniger coupling \(\gamma_0=c/\rho_0\) are displayed on logarithmic scales. The numerical results follow the conjectured interaction dependence across the full range shown, with a systematic relative difference of approximately \(3\%\)--\(5\%\).}
    \label{fig:conjecture-comparison}
\end{figure}

For the uniform ground state at density $\rho_0$, the occupied Bethe
rapidities form the interval $[-q,q]$, where $q$ is the Fermi rapidity. Its
relation to $c$ and $\rho_0$ is determined by the Lieb equation \eqref{eq:Lieb-equation} reviewed in
Sec.~\ref{sec:hydro}. We define the dimensionless ratio $r=c/q$.
Based on the asymptotic analysis of the Fredholm-determinant representation
presented in Sec.~\ref{sec:conjecture}, we propose the following conjecture
for the emptiness formation probability:
\begin{equation}\label{eq:P(r)}
    P(R)
    =\exp\left\{
    -\frac{(qR)^2}{2}\int_{-1}^{1}t g(t)\,dt
    +o(R^2)\right\},
\end{equation}
where $g(t)$ is determined by the linear integral equation
\begin{equation}\label{eq:dimless-int-eq}
    \int_{-1}^{1}g(s)
    \log\left(1+\frac{r^2}{(t-s)^2}\right)ds
    =4t.
\end{equation}
The rate function in the physical normalization of
Eq.~\eqref{eq:rate-function} is therefore
\begin{equation}
    f(\gamma_0)
    =\frac{1}{2}\left(\frac{q}{\rho_0}\right)^2
    \int_{-1}^{1}t g(t)\,dt
    =\frac{1}{2}\left(\frac{\gamma_0}{r}\right)^2
    \int_{-1}^{1}t g(t)\,dt.
    \label{eq:physical-rate-from-g}
\end{equation}
For a prescribed physical coupling $\gamma_0=c/\rho_0$, the corresponding
$r=c/q$ is fixed by solving the ground-state Lieb equation \eqref{eq:Lieb-equation} and imposing
$\rho_0=\int_{-q}^{q}\rho_p(\lambda)\,d\lambda$. Together with Eqs.~\eqref{eq:dimless-int-eq} and
\eqref{eq:physical-rate-from-g}, these relations determine \(f(\gamma_0)\)
without fitted parameters. 

\begin{figure}[ht!]
    \centering
    \includegraphics[width=0.45\linewidth, trim=0 0 80pt 0, clip]{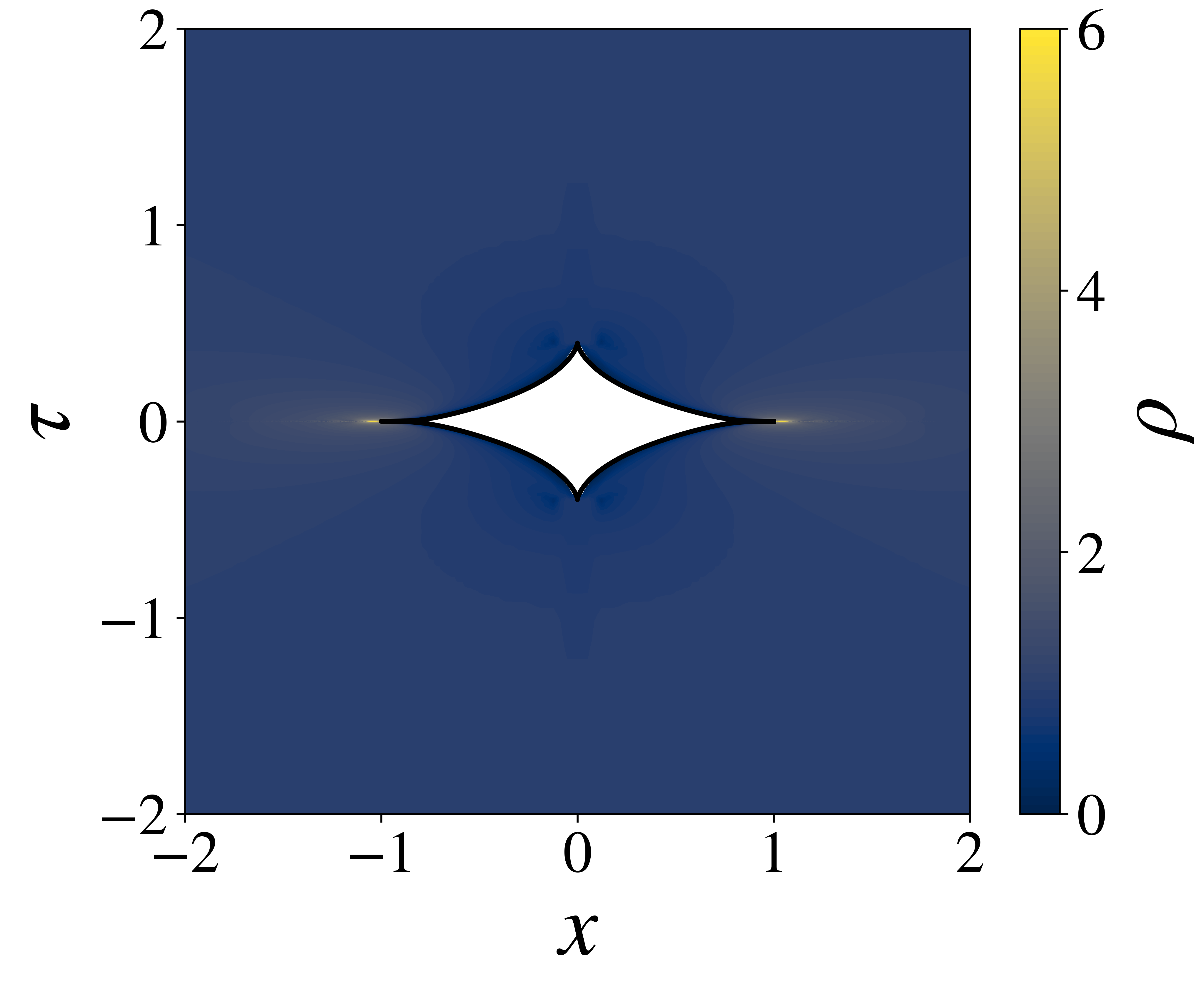}
    \includegraphics[width=0.535\linewidth]{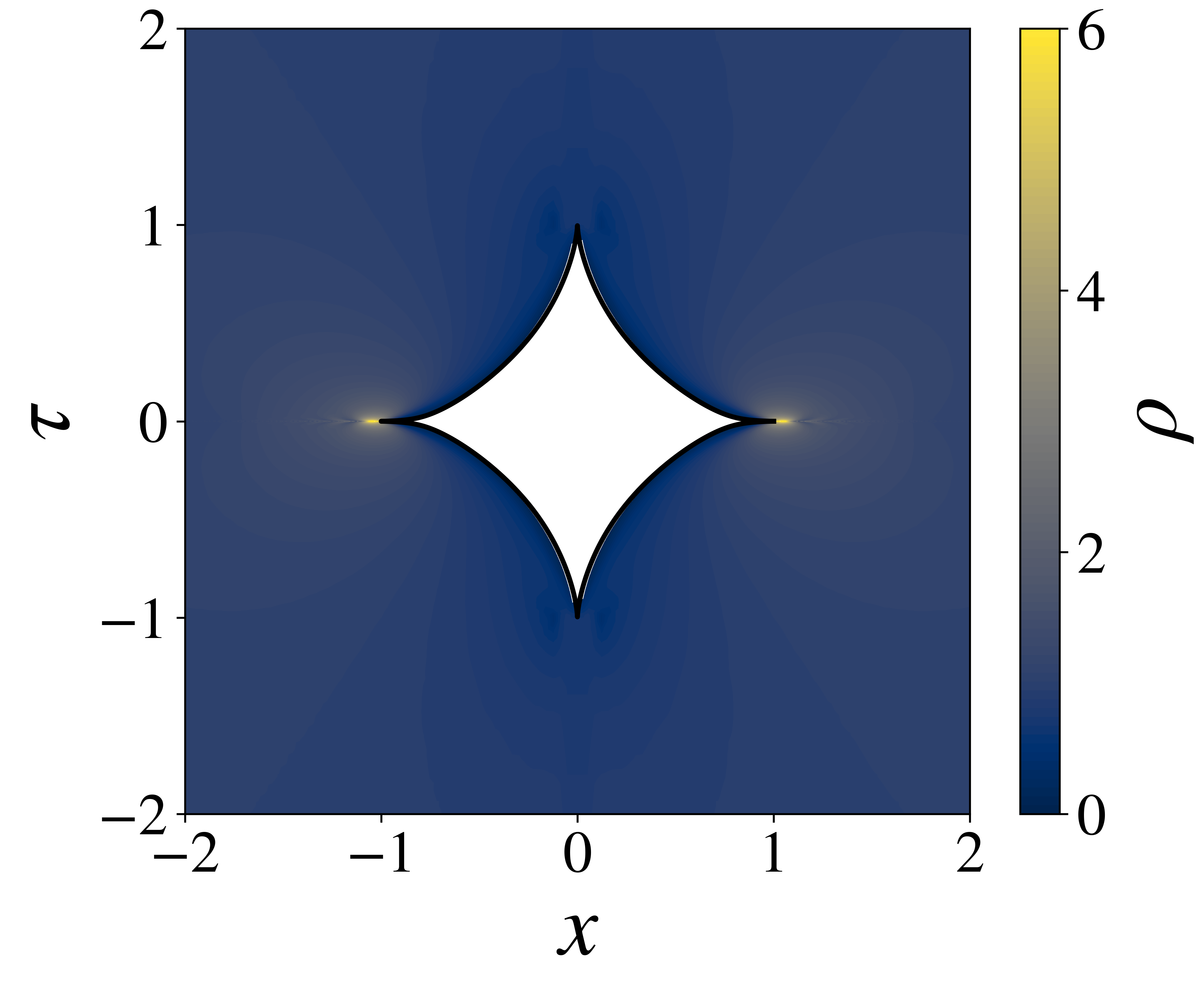}
    \caption{Representative numerical emptiness instantons at strong coupling, \(\gamma_0=20\) (left), and intermediate coupling, \(\gamma_0=1\) (right). The color scale shows the density \(\rho(x,\tau)\). The axes show the dimensionless coordinates \(x/R\) and \(\hbar\rho_0\tau/(mR)\); the numerical calculations use \(R=\rho_0=\hbar=m=1\). The empty interval \(|x|<R\) is formed at \(\tau=0\), and its astroid-like boundary closes at the critical imaginary times \(\tau=\pm\tau_c\).}
    \label{fig:instanton-examples}
\end{figure}

We test this conjecture independently using the hydrodynamic formulation. For various values of $\gamma_0$, we numerically minimize the Euclidean action subject to the formation of an empty region at $\tau = 0$. Two examples of the resulting emptiness instantons are shown in Fig.~\ref{fig:instanton-examples}. The boundary of the empty region has an astroid-like shape, as previously found for free fermions and, more generally, for polytropic fluids \cite{abanov-hydro,abanov2025polytropicEFP}.

We find good agreement between the numerical results and the conjectured EFP, as shown in Fig.~\ref{fig:conjecture-comparison}. In addition, the conjecture reproduces the known leading weak- and strong-coupling limits, while its first subleading corrections agree with those obtained independently from hydrodynamics. 

This paper is organized as follows. In Sec.~\ref{sec:conjecture}, we review the exact KIW dual-field representation, formulate the required uniform large-distance asymptotic hypothesis, and show that the resulting joint dual-field saddle leads to our conjectured integral equation. In Sec.~\ref{sec:hydro}, we review the hydrodynamic formulation of the EFP and the Lieb–Liniger equation of state. Section~\ref{sec:results} presents independent weak- and strong-coupling checks and compares the conjecture with numerical hydrodynamic instantons. We conclude in Sec.~\ref{sec:discussion} with a discussion of the scope of the conjecture and possible extensions. Technical details of the numerical method and the weak- and strong-coupling calculations are collected in the appendices.

%%%%%%%%%%%%%%
\section{Fredholm-determinant formulation and conjecture}
\label{sec:conjecture}

%%%%%%%%%%
\subsection{Exact dual-field representation}

Given the nontrivial form of the Bethe wave function, it is useful to
represent the EFP in terms of an integrable Fredholm operator. In the
thermodynamic limit, the result of Korepin, Its, and Waldron (KIW) reads
\cite{korepin1995probabilityphaseseparationbose,korepin1997quantum}
\begin{equation}
    P(x)=
    \frac{\bra 0\det(I+\hat V_c)\ket 0}
    {\det(I-\hat\Delta/2\pi)}.
    \label{eq:exact-fredholm-representation}
\end{equation}
Here \(x=2R\) is the length of the empty interval, and \(\hat\Delta\) is defined by the Lieb--Liniger
scattering kernel
\begin{equation}
    \Delta(\lambda,\mu)
    =\frac{2c}{(\lambda-\mu)^2+c^2}.
    \label{eq:LL-scattering-kernel}
\end{equation}
The denominator is therefore independent of \(x\) and does not contribute
to the leading large-\(x\) exponent. The operator in the numerator acts as
\begin{equation}
    (\hat V_c u)(\lambda)
    =\int_{-q}^{q}V_c(\lambda,\mu)u(\mu)\,d\mu.
\end{equation}
The kernel \(V_c(\lambda,\mu)\) is conveniently expressed in terms of the
function
\begin{equation}
    E(\lambda,\mu)
    =
    \exp\left[
        \frac{ix(\lambda-\mu)}{2}
        +\frac{\mathring\varphi(\lambda)-\mathring\varphi(\mu)}{2}
    \right],
    \qquad
    E(\mu,\lambda)=E(\lambda,\mu)^{-1}.
\end{equation}
In terms of \(E\),
\begin{equation}
    V_c(\lambda,\mu)
    =
    -\frac{c}{2\pi(\lambda-\mu)}
    \left[
        \frac{E(\lambda,\mu)}{\lambda-\mu+ic}
        +
        \frac{E(\mu,\lambda)}{\lambda-\mu-ic}
    \right].
    \label{eq:exact-Vc}
\end{equation}
The dual field is decomposed as
\begin{equation}
    \mathring\varphi(\lambda)
    =\mathring p(\lambda)+\mathring q(\lambda),
\end{equation}
where \(\mathring p\) and \(\mathring q\) satisfy the commutation relations
and vacuum conditions
\begin{align}
    [\mathring p(\lambda),\mathring q(\mu)]
    &=C_c(\lambda,\mu)
    \equiv
    \log\frac{c^2}{c^2+(\lambda-\mu)^2},
    \label{eq:dual-covariance}\\
    [\mathring p(\lambda),\mathring p(\mu)]
    &=[\mathring q(\lambda),\mathring q(\mu)]=0,
    \qquad \qquad \qquad
    \mathring p(\lambda)\ket0=0=\bra0\mathring q(\lambda).
\end{align}
Since $C_c(\lambda,\mu)$ is symmetric,
$[\mathring\varphi(\lambda),\mathring\varphi(\mu)]=0$. The dual fields
may therefore be treated as commuting functional variables until the vacuum
expectation value is taken.

For a \(c\)-number source \(J\), the auxiliary vacuum average is
\begin{equation}
    \left\langle0\left|
    \exp\left\{\int_{-q}^{q}J(\lambda)
    \mathring\varphi(\lambda)\,d\lambda\right\}
    \right|0\right\rangle
    =
    \exp\left\{
    \frac12\iint_{-q}^{q}
    J(\lambda)C_c(\lambda,\mu)J(\mu)
    \,d\lambda\,d\mu
    \right\}.
    \label{eq:dual-gaussian-functional}
\end{equation}
Thus, the auxiliary vacuum averaging can be represented formally as a
Gaussian functional integration with covariance kernel \(C_c\). All occurrences
of the dual field in Eq.~\eqref{eq:exact-Vc} are through differences
\(\mathring\varphi(\lambda)-\mathring\varphi(\mu)\). The corresponding
sources are therefore neutral, \(\int J=0\), and the constant mode of the
field decouples. This Gaussian representation can be applied term by term
in the Fredholm expansion and also underlies the normal-ordering approach
to dual-field averages developed in
Refs.~\cite{KorepinSlavnov1997,Slavnov1998Fredholm}.

Up to this point, the construction is precisely the exact KIW
dual-field representation. The difference begins only when the large-$x$
asymptotics of the determinant is combined with the dual-field vacuum
expectation. A contribution that is subleading for a fixed dual field need
not remain subleading under this average.

%%%%%%%%%%
\subsection{Point of departure from the KIW analysis}
\label{sec:KIW-departure}

Starting from the exact representation Eq.~\eqref{eq:exact-fredholm-representation}, KIW proposed the following
large-$x$ asymptotic formula for the determinant at fixed dual field:
\begin{equation}
    \log\det(I+\hat V_c[\mathring\varphi])
    =-\frac{(xq)^2}{8}
    +x\,\mathcal A_{\rm KIW}[\mathring\varphi]
    +o(x),
    \label{eq:KIW-fixed-field-asymptotic}
\end{equation}
where $\mathcal A_{\rm KIW}$ is linear in $\mathring\varphi$. They explicitly referred to both the determinant asymptotic and the resulting EFP formula as conjectures, with analyticity and large-spectral-parameter decay assumptions on an auxiliary Riemann--Hilbert quantity serving as their main hypotheses
\cite{korepin1995probabilityphaseseparationbose}. 

The term linear in $\mathring\varphi$ in
Eq.~\eqref{eq:KIW-fixed-field-asymptotic} acts as a source of order $x$ in
the Gaussian dual-field average. We therefore expect the dominant fields to
scale as $\mathring\varphi=O(x)$ and introduce the rescaled field
\begin{equation}
    \mathring\varphi(\lambda)=x\Phi(\lambda),
    \qquad \Phi(\lambda)=O(1)
    \quad (x\to\infty).
    \label{eq:rescaled-dual-field}
\end{equation}

At fixed \(\mathring\varphi=O(1)\),
Eq.~\eqref{eq:KIW-fixed-field-asymptotic} determines the terms of order
\(x^2\) and \(x\), but leaves the \(o(x)\) remainder undetermined. This
remainder can contain terms quadratic in \(\mathring\varphi\), which are
\(O(1)\) at fixed field but become \(O(x^2)\) under the scaling in
Eq.~\eqref{eq:rescaled-dual-field}. Such terms therefore cannot be
discarded before taking the dual-field vacuum average. To retain the
relevant quadratic contribution, introduce
\begin{equation}
    K_0(\lambda,\mu)
    =\log\frac{c^2}{(\lambda-\mu)^2},
\end{equation}
and use the notation
\begin{equation}
    \langle u,Av\rangle
    =
    \iint_{-q}^{q}
    u(\lambda)A(\lambda,\mu)v(\mu)
    \,d\lambda\,d\mu .
\end{equation}
In expressions such as \(K_0^{-1}\lambda\), \(\lambda\) denotes the
function \(f(\mu)=\mu\); explicitly,
\[
    (K_0^{-1}\lambda)(\nu)
    =
    \int_{-q}^{q}K_0^{-1}(\nu,\mu)\,\mu\,d\mu .
\]
We formulate the double-scaled determinant hypothesis as
\begin{equation}
    \log\det(I+\hat V_c[x\Phi])
    =
    \frac{x^2}{2}
    \left\langle
    \Phi+i\lambda,K_0^{-1}(\Phi+i\lambda)
    \right\rangle+o(x^2),
    \label{eq:uniform-det-hypothesis}
\end{equation}
uniformly for $\Phi$ in a neighborhood of the saddle. In terms of the
unscaled field $\mathring\varphi=x\Phi$, the quadratic form on the right-hand
side expands as
\begin{align}
    \frac12\left\langle
    \mathring\varphi+ix\lambda,
    K_0^{-1}(\mathring\varphi+ix\lambda)
    \right\rangle
    ={}&-\frac{x^2}{2}\langle\lambda,K_0^{-1}\lambda\rangle
    +ix\langle\mathring\varphi,K_0^{-1}\lambda\rangle
    +\frac12\langle\mathring\varphi,
    K_0^{-1}\mathring\varphi\rangle.
    \label{eq:quadratic-dual-field-completion}
\end{align}
The first two terms reproduce the field-independent and linear-in-field
terms in the KIW fixed-field asymptotics. The last term is \(O(1)\) at
fixed \(\mathring\varphi\), but becomes \(O(x^2)\) for
\(\mathring\varphi=x\Phi\). It therefore modifies the leading EFP exponent
and marks the point where our proposed asymptotic completion departs from
Eq.~\eqref{eq:KIW-fixed-field-asymptotic}.

A one-dimensional Gaussian integral illustrates the nonuniformity:
\begin{equation}
    I(x)
    =\int_{-\infty}^{\infty}d\phi\,
    \exp\left[
        -\frac{\phi^2}{2}
        +x\phi
        +\frac{b\phi^2}{2}
    \right]
    \propto
    \exp\left[\frac{x^2}{2(1-b)}\right].
    \label{eq:scalar-gaussian-analogy}
\end{equation}
At fixed \(\phi\), the \(b\)-term is \(O(1)\), whereas the \(x\phi\)-term
is \(O(x)\). Nevertheless, the \(b\)-term combines with the original
Gaussian quadratic term and changes the coefficient of \(x^2\) in
\(\log I(x)\). The quadratic dual-field contribution plays the analogous
role in the dual-field vacuum average.

Hence, we combine the determinant limit with the dual-field average.
Up to factors independent of \(x\), the Gaussian representation
Eq.~\eqref{eq:dual-gaussian-functional} gives schematically
\begin{equation}
    P(x)\propto
    \int\mathcal D\mathring\varphi\,
    \exp\left\{
        \log\det(I+\hat V_c[\mathring\varphi])
        -\frac12
        \langle\mathring\varphi,
        C_c^{-1}\mathring\varphi\rangle
    \right\}.
    \label{eq:dual-functional-integral}
\end{equation}
Using the rescaling in Eq.~\eqref{eq:rescaled-dual-field}, define the
determinant rate functional
\begin{equation}
    \mathcal D_c[\Phi]
    =
    \lim_{x\to\infty}\frac{1}{x^2}
    \log\det(I+\hat V_c[x\Phi]).
    \label{eq:det-rate-functional}
\end{equation}
A saddle-point evaluation of
Eq.~\eqref{eq:dual-functional-integral} then yields
\begin{equation}
    \frac{1}{x^2}\log P(x)
    =
    \mathcal D_c[\Phi_*]
    -\frac12\langle\Phi_*,C_c^{-1}\Phi_*\rangle
    +o(1),
    \label{eq:joint-dual-saddle}
\end{equation}
where \(\Phi_*\) satisfies
\begin{equation}
    \left.
    \frac{\delta}{\delta\Phi}
    \left[
        \mathcal D_c[\Phi]
        -\frac12\langle\Phi,C_c^{-1}\Phi\rangle
    \right]
    \right|_{\Phi=\Phi_*}
    =0.
\end{equation}
The inverse \(C_c^{-1}\) is understood on neutral fields, and the relevant
saddle \(\Phi_*\) may in general be complex.

To relate Eq.~\eqref{eq:uniform-det-hypothesis} to known asymptotic results
for generalized sine kernels, it is useful to decompose the exact kernel.
In terms of the function \(E(\lambda,\mu)\) introduced above, one has
\begin{equation}
    V_c(\lambda,\mu)
    =
    \frac{i}{2\pi}
    \frac{E(\lambda,\mu)-E(\mu,\lambda)}{\lambda-\mu}
    -\frac{i}{2\pi}
    \frac{E(\lambda,\mu)}{\lambda-\mu+ic}
    +\frac{i}{2\pi}
    \frac{E(\mu,\lambda)}{\lambda-\mu-ic}.
    \label{eq:Vc-gsk-decomposition}
\end{equation}
The first term is precisely the generalized sine kernel studied by
Gharakhloo, Its, and Kozlowski, with large parameter \(x\), phase \(p(\lambda)=\lambda\), and, in the
notation of Ref.~\cite{Gharakhloo2019}, external field
\(g(\lambda)=\mathring\varphi(\lambda)\). Using a Riemann--Hilbert analysis, they derived the
large-\(x\) asymptotics of its Fredholm determinant when the phase and
external field are fixed analytic functions. Motivated by applications to
the EFP away from the free-fermion point, they also explicitly formulated
the two-step procedure of first obtaining the determinant asymptotics at
fixed dual field and then performing the dual-field vacuum average.

The present problem requires an extension of their result in two respects.
First, the vacuum average is dominated by fields
\(\mathring\varphi=x\Phi\), outside the fixed-field regime considered in
Ref.~\cite{Gharakhloo2019}. Second, the exact Lieb--Liniger kernel contains
the two shifted finite-\(c\) terms in
Eq.~\eqref{eq:Vc-gsk-decomposition}. Our hypothesis assumes that the
generalized sine-kernel asymptotics extends uniformly to
\(\mathring\varphi=x\Phi\) near the joint saddle and that the shifted terms
do not change the leading \(x^2\) rate functional. This double-scaled
statement remains unproved. Accordingly, we subject the resulting
conjecture to independent weak- and strong-coupling tests and to comparison
with numerical hydrodynamic calculations in Sec.~\ref{sec:results}.

%%%%%%%%%%%
\subsection{Evaluation of the joint saddle}

Under the determinant hypothesis
Eq.~\eqref{eq:uniform-det-hypothesis}, the functional whose saddle
determines Eq.~\eqref{eq:joint-dual-saddle} is
\begin{equation}
    \mathcal S[\Phi]
    =-\frac12\langle\Phi,C_c^{-1}\Phi\rangle
    +\frac12\left\langle
    \Phi+i\lambda,K_0^{-1}(\Phi+i\lambda)
    \right\rangle.
    \label{eq:joint-action-Phi}
\end{equation}
Introduce the difference of the two kernels,
\begin{equation}
    L_c(\lambda,\mu)
    \equiv K_0(\lambda,\mu)-C_c(\lambda,\mu)
    =
    \log\left(1+\frac{c^2}{(\lambda-\mu)^2}\right).
    \label{eq:Lc-kernel}
\end{equation}
Varying Eq.~\eqref{eq:joint-action-Phi} with respect to \(\Phi\) and
solving the resulting linear equation gives
\begin{equation}
    \Phi_*
    =
    C_cL_c^{-1}(i\lambda).
    \label{eq:Phi-saddle}
\end{equation}
The action evaluated at this saddle is
\begin{equation}
    \mathcal S[\Phi_*]
    =
    \frac12\langle i\lambda,L_c^{-1}i\lambda\rangle
    =
    -\frac12\langle\lambda,L_c^{-1}\lambda\rangle.
    \label{eq:on-shell-joint-action}
\end{equation}
Define the function
\begin{equation}
    \mathfrak g=4L_c^{-1}\lambda.
    \label{eq:g-operator-definition}
\end{equation}
Equivalently, \(\mathfrak g\) satisfies
\begin{equation}
    \int_{-q}^{q}\mathfrak g(\mu)
    \log\left(1+\frac{c^2}{(\lambda-\mu)^2}\right)d\mu
    =4\lambda.
    \label{eq:integral-eq-rate}
\end{equation}
In terms of \(\mathfrak g\), the on-shell action becomes
\begin{equation}
    \mathcal S[\Phi_*]
    =
    -\frac18\int_{-q}^{q}
    \lambda\mathfrak g(\lambda)\,d\lambda.
\end{equation}
Combining the on-shell action with
Eq.~\eqref{eq:joint-dual-saddle}, the conjectured leading EFP is
\begin{equation}
    P(x)
    =\exp\left\{
    -\frac{x^2}{8}\int_{-q}^{q}
    \lambda\mathfrak g(\lambda)\,d\lambda+o(x^2)
    \right\}.
    \label{eq:integral-eq-f}
\end{equation}

To establish uniqueness of the solution and verify that the decay rate in
Eq.~\eqref{eq:integral-eq-f} is positive, it is convenient to consider the
Fourier transform of the kernel on the full line. Since \(L_c(\lambda,\mu)\) depends only on the difference
\(u=\lambda-\mu\), write
\[
    L_c(u)=\log\left(1+\frac{c^2}{u^2}\right)
\]
and define
\[
    \widehat L_c(k)
    =\int_{-\infty}^{\infty}e^{-iku}L_c(u)\,du .
\]
Using
\begin{equation}
    L_c(u)
    =\int_0^c\frac{2\kappa}{u^2+\kappa^2}\,d\kappa,
    \qquad
    \widehat L_c(k)
    =2\pi\frac{1-e^{-c|k|}}{|k|}>0,
    \label{eq:Lc-positive}
\end{equation}
with \(\widehat L_c(0)=2\pi c\), we conclude that \(L_c\) defines a
strictly positive quadratic form. Thus, any solution of
Eq.~\eqref{eq:integral-eq-rate} is unique. Reflection symmetry implies that
\(\mathfrak g\) is odd, and
\begin{equation}
    \int_{-q}^{q}\lambda\mathfrak g(\lambda)\,d\lambda
    =\frac14\langle\mathfrak g,L_c\mathfrak g\rangle>0,
\end{equation}
so the rate in Eq.~\eqref{eq:integral-eq-f} has the required sign.

Finally, after rescaling
\begin{equation*}
    x=2R,\qquad \lambda=qt,\qquad \mu=qs,\qquad
    r=\frac{c}{q},\qquad g(t)=\mathfrak g(qt),
\end{equation*}
Eqs.~\eqref{eq:integral-eq-rate} and \eqref{eq:integral-eq-f}
reduce precisely to the conjecture stated in
Eqs.~\eqref{eq:P(r)} and \eqref{eq:dimless-int-eq}.

%%%%%%%%%
\subsection{Equivalent scattering-kernel form}

It is useful to rewrite the result in terms of an antiderivative of the
solution. Define
\begin{equation}
    \mathcal G(\lambda)
    =\int_{-q}^{\lambda}\mathfrak g(\mu)\,d\mu.
\end{equation}
Because $\mathfrak g$ is odd, $\mathcal G(-q)=\mathcal G(q)=0$, and
integration by parts gives
\begin{equation}
    \int_{-q}^{q}\lambda\mathfrak g(\lambda)\,d\lambda
    =-\int_{-q}^{q}\mathcal G(\lambda)\,d\lambda.
\end{equation}
Substituting $\mathfrak g=\mathcal G'$ into
Eq.~\eqref{eq:integral-eq-rate} and integrating by parts once more yields
\begin{equation}
    \mathrm{P.V.}\int_{-q}^{q}
    \frac{2c^2\,\mathcal G(\mu)}
    {(\lambda-\mu)[(\lambda-\mu)^2+c^2]}
    \,d\mu=-4\lambda.
\end{equation}
In terms of the Lieb--Liniger scattering kernel
\eqref{eq:LL-scattering-kernel}, this becomes
\begin{equation}
    \mathrm{P.V.}\int_{-q}^{q}
    \frac{\Delta(\lambda,\mu)}{\lambda-\mu}
    \mathcal G(\mu)\,d\mu
    =-\frac{4\lambda}{c},
    \qquad \mathcal G(\pm q)=0.
    \label{eq:first-form}
\end{equation}
This form makes explicit the direct relation between the conjectured EFP
rate function and the two-body Lieb--Liniger scattering kernel.

%%%%%%%%%%%%%%%%%%%%%%%
\section{Emptiness formation probability from hydrodynamics} \label{sec:hydro}
\begin{figure}[t!]
    \centering
    \includegraphics[width=0.6\linewidth]{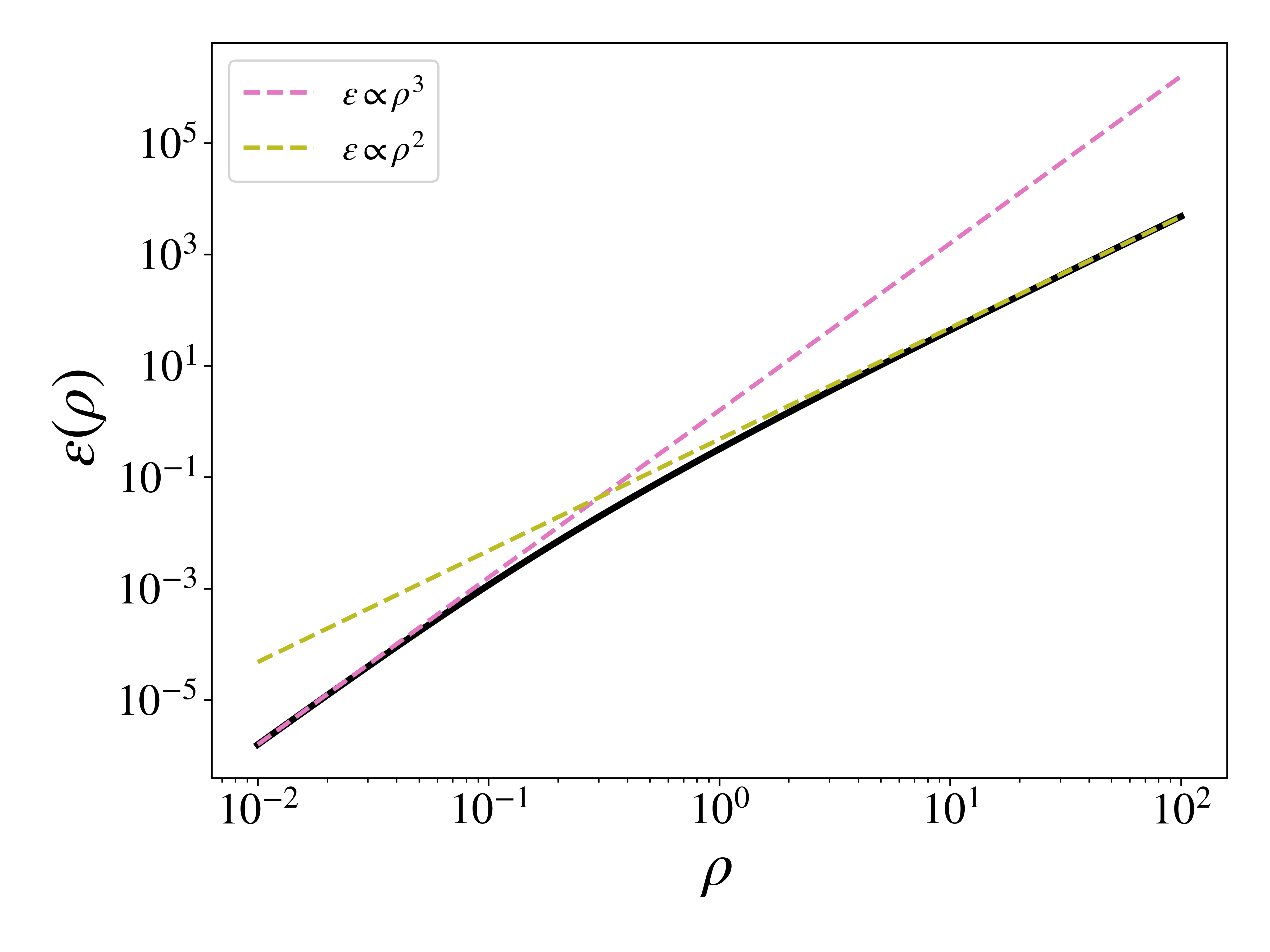}
    \caption{Lieb--Liniger ground-state energy density \(\epsilon(\rho)\) at fixed coupling \(c=1\) (solid black line). At low density, \(\gamma=c/\rho\gg1\), and the equation of state approaches the Tonks--Girardeau asymptote \(\epsilon(\rho)\simeq \pi^2\rho^3/6\) (magenta dashed line). At high density, \(\gamma\ll1\), it approaches the Gross--Pitaevskii asymptote \(\epsilon(\rho)\simeq c\rho^2/2\) (yellow dashed line). Both axes are logarithmic.}
    \label{fig:LL-eos}
\end{figure}

At leading order in the size of a macroscopic empty interval, the EFP can
be formulated as a variational problem for the Euclidean hydrodynamic
action \cite{abanov-hydro}. We use the
background-subtracted action
\begin{equation}\label{eq:hydro-action}
    S_E[\rho,v]
    =\int d\tau\,dx\left[
       \frac{\rho v^2}{2}+h(\rho)\right],
    \qquad
    h(\rho)=\epsilon(\rho)-\epsilon(\rho_0)
    -\mu_0(\rho-\rho_0),
\end{equation}
where $\mu_0=\epsilon'(\rho_0)$ and $v$ denotes the real velocity field on
the Euclidean saddle. The subtraction makes the uniform state
$(\rho,v)=(\rho_0,0)$ contribute zero action. The density and velocity obey
the continuity equation
\begin{equation}
    \partial_\tau\rho+\partial_x(\rho v)=0.
    \label{eq:euclidean-continuity}
\end{equation}
The only model-specific input is the ground-state energy density
$\epsilon(\rho)$. For the Lieb--Liniger gas, the ground-state rapidity density
\(\rho_p(\lambda)\) satisfies the Lieb equation
\cite{lieb1963exact,korepin1997quantum},
\begin{equation}
    2\pi \rho_p(\lambda) = 1 + \int_{-q}^q d\mu\, \Delta(\lambda,\mu)\, \rho_p(\mu),
 \label{eq:Lieb-equation}
\end{equation}
where $q$ is the Fermi rapidity, determined implicitly by the particle density,
\begin{equation}
    \rho = \int_{-q}^q d\lambda\, \rho_p(\lambda).
\end{equation}
The ground-state energy density is then
\begin{equation}
    \epsilon(\rho) = \int_{-q}^q d\lambda\,\frac{\lambda^2}{2}\rho_p(\lambda). 
\end{equation}
As an illustration, Fig.~\ref{fig:LL-eos} shows the equation of state
\(\epsilon(\rho)\) at fixed \(c=1\). As the density decreases, the ratio
\(c/\rho\) increases and the equation of state approaches the
Tonks--Girardeau form \(\epsilon(\rho)\propto\rho^3\). At high density, it
approaches the Gross--Pitaevskii form
\(\epsilon(\rho)\propto\rho^2\).

It is convenient to enforce Eq.~\eqref{eq:euclidean-continuity} by a
displacement field $u(x,\tau)$ such that
\begin{equation}
    \rho = \partial_x u, \qquad \rho v = -\partial_\tau u.
\end{equation}
The admissible fields must obey the EFP boundary conditions, 
\begin{equation}
    \rho(x,0)=0\quad (|x|<R),
    \qquad
    (\rho,v)\to(\rho_0,0)
    \quad\text{as } |x|\to\infty
    \text{ or } |\tau|\to\infty.
    \label{eq:EFP-hydro-boundary}
\end{equation}
The leading large-$R$ probability is determined by the minimum of
Eq.~\eqref{eq:hydro-action} subject to these conditions,
\begin{equation}
    P(R)\asymp \exp[-S_E^{\mathrm{opt}}],
    \qquad
    S_E^{\mathrm{opt}}=\min S_E[\rho,v].
    \label{eq:EFP-Euclidean-saddle}
\end{equation}
This is a long-wavelength hydrodynamic description in which gradient
corrections to the local energy functional are omitted. For an interval much larger than the microscopic correlation length, such
corrections are subleading to the \(R^2\) term determined by
Eq.~\eqref{eq:EFP-Euclidean-saddle}.

The minimizer is the emptiness instanton, describing the optimal rare
fluctuation in which the interval \([-R,R]\) contains no particles at
\(\tau=0\). In imaginary time, the empty region exists for \(|\tau|<\tau_c\) and closes
at \(\tau=\pm\tau_c\), forming an astroid-like boundary
\cite{abanov-hydro,yeh_emptiness_2020,yeh2022emptiness,
abanov2025polytropicEFP}. Restoring units, the critical time has the scaling
form
\begin{equation}
    \tau_c(\gamma_0)
    =
    \frac{mR}{\hbar\rho_0}\mathcal T(\gamma_0).
\end{equation}

For the symmetric interval, parity and time-reversal symmetry allow the
Euclidean action to be minimized in the first quadrant of the
\((x,\tau)\) plane. We parameterize the boundary by
\begin{equation}
    \tau=\tau_c\zeta^3,
    \qquad
    x=X(\zeta),
    \qquad
    X(0)=R,\quad X(1)=0.
\end{equation}
The cubic parametrization smooths the expected astroid-like behavior near
\(\tau=0\) and, for a uniform grid in \(\zeta\), concentrates the temporal
resolution in that region. Introducing the transverse coordinate \(y\geq0\) as
\begin{equation}
    x=X(\zeta)+y
\end{equation}
fixes the instanton boundary at \(y=0\). Further details of the numerical
implementation are given in Appendix~\ref{app:numerical-strat}.

%%%%%%%%%%%%%
\section{Results and comparisons}\label{sec:results}

%%%%%%%%%%%%%
\subsection{Weak-coupling comparison}

As \(\gamma_0\to0\), the leading mean-field equation of state is polytropic, \(\epsilon(\rho)\propto\rho^{\gamma_{\rm pol}}\), with \(\gamma_{\rm pol}=2\). The exact hydrodynamic instanton for this equation
of state, obtained in Ref.~\cite{abanov2025polytropicEFP}, yields
\begin{equation}
    f(\gamma_0)\sim\frac{16}{3\pi}\sqrt{\gamma_0}.
\end{equation}
First-order perturbation theory around this instanton gives the subleading
term,
\begin{equation}
    f(\gamma_0)
    =\frac{16}{3\pi}\sqrt{\gamma_0}
    +\frac{4(1-\log 4)}{\pi^2}\gamma_0
    +O(\gamma_0^{3/2}).
    \label{eq:weak-hydro-result-main}
\end{equation}
Direct numerical quadrature of the first-order on-shell hydrodynamic action gives \(A_1^{\rm num}\simeq-0.15656\), in agreement with the analytical value \(A_1=4(1-\log 4)/\pi^2\simeq-0.15655\) in Eq.~\eqref{eq:weak-hydro-result-main}.
The perturbative calculation and the exact evaluation of its coefficient are
given in Appendices~\ref{app:perturb} and~\ref{app:weak-integral}.

The same expansion follows from the conjecture. For $r=c/q\to0$, its kernel
has the distributional limit
\begin{equation}
    \log\left(1+\frac{r^2}{(t-s)^2}\right)
    \sim 2\pi r\,\delta(t-s),
\end{equation}
so that $g(t)\sim2t/(\pi r)$. Together with the weak-coupling relation
between $r$ and $\gamma_0$, Eq.~\eqref{eq:physical-rate-from-g} reproduces
the leading term in Eq.~\eqref{eq:weak-hydro-result-main}. The next order is
singular because the expansion is nonuniform near $t=\pm1$. Matched
asymptotics and Wiener--Hopf factorization give
\cite{santana2026weakcouplinglimitlatticenonlinear,Tracy2016}\footnote{In
terms of the Lieb--Liniger Luttinger parameter
$K=\pi\rho_0/v_s$, the weak-coupling equation of state gives
$K=\pi/\sqrt{\gamma_0}+1/4+O(\sqrt{\gamma_0})$. Consequently, the same result takes the form
\(f=16/(3K)+4(4/3-\log 4)/K^2+o(K^{-2})\).}
\begin{equation}
    f(\gamma_0)
    =\frac{16}{3\pi}\sqrt{\gamma_0}
    +\frac{4(1-\log 4)}{\pi^2}\gamma_0
    +o(\gamma_0).
\end{equation}
Thus, the conjecture reproduces both the leading weak-coupling behavior
and the first nontrivial correction obtained independently from
hydrodynamics. Details of the conjecture expansion are given in
Appendix~\ref{app:expansion}.

%%%%%%%%%
\subsection{Strong-coupling comparison}

As $\gamma_0\to\infty$, the Lieb--Liniger gas approaches the
Tonks--Girardeau gas and the leading EFP rate is the free-fermion result
$f=\pi^2/2$. The first hydrodynamic correction, obtained from the
expanded-volume mapping described in Appendix~\ref{app:perturb}, is
\begin{equation}
    f(\gamma_0)
    =\frac{\pi^2}{2}\left(1-\frac{4}{\gamma_0}
    +O(\gamma_0^{-2})\right).
    \label{eq:strong-hydro-result-main}
\end{equation}
Expanding the conjecture gives
\begin{equation}
    f(\gamma_0)
    =\frac{\pi^2}{2}\left[
       1-\frac{4}{\gamma_0}
       +\frac{12+\pi^2/2}{\gamma_0^2}
       +O(\gamma_0^{-3})
    \right].
    \label{eq:strong-conjecture-result-main}
\end{equation}
Thus the conjecture agrees with the independent result through order
$1/\gamma_0$. The coefficient of $1/\gamma_0^2$ in
Eq.~\eqref{eq:strong-conjecture-result-main} is a further prediction that is
not independently checked here.

\subsection{Numerical results}

\begin{figure}
    \centering
    \includegraphics[width=0.49\linewidth]{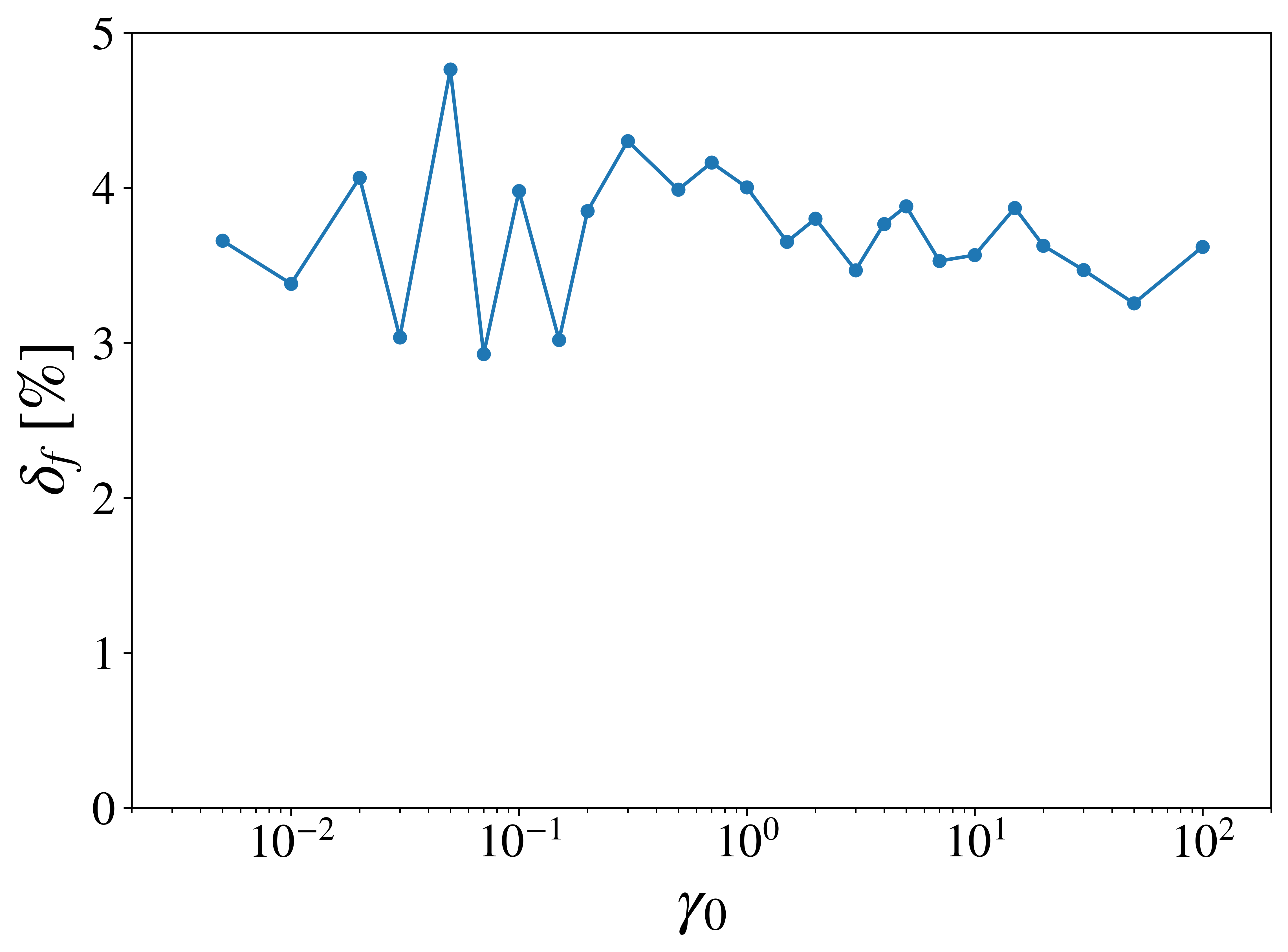}
    \includegraphics[width=0.49\linewidth]{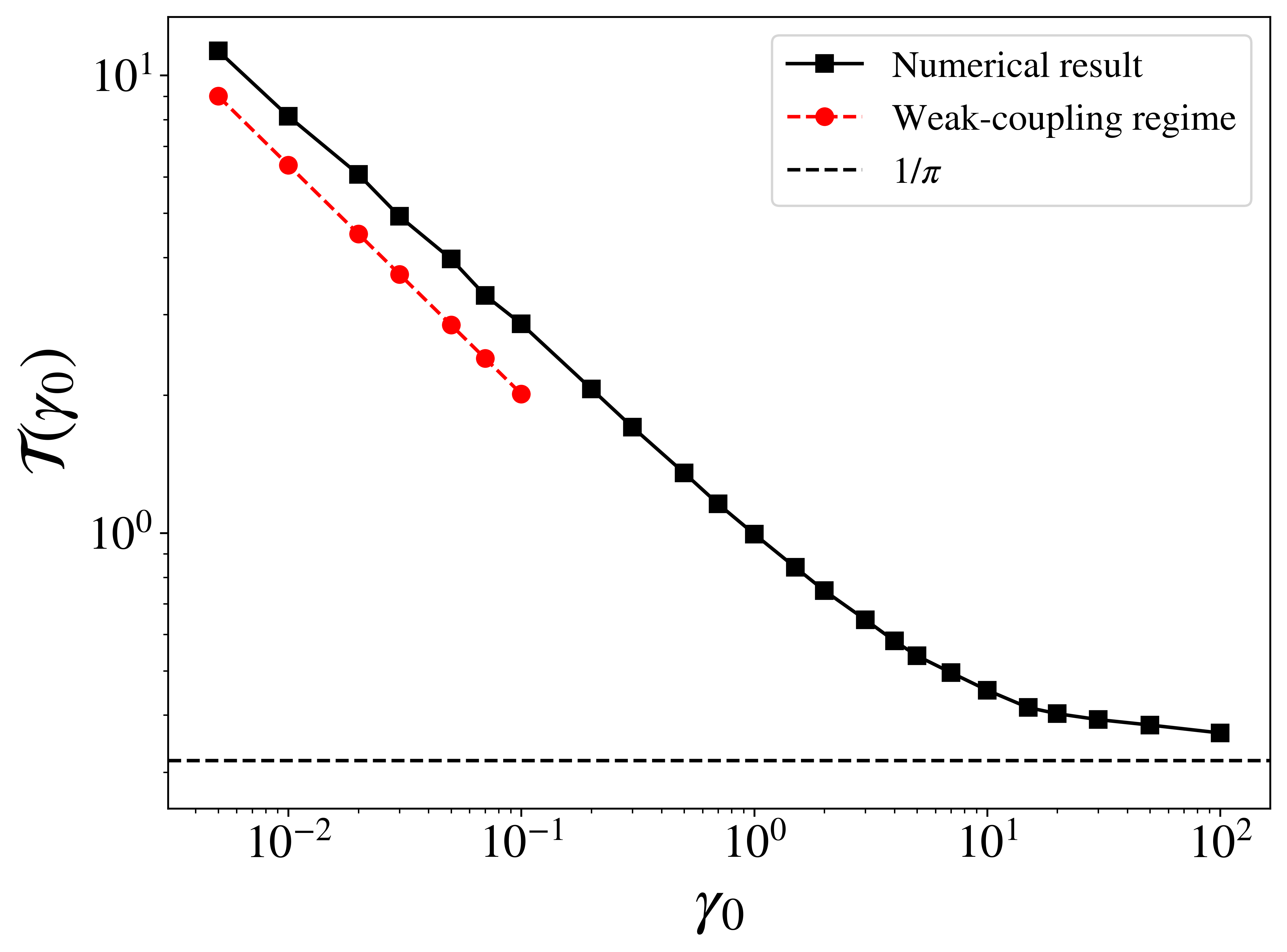}
    \caption{Numerical comparison across the interaction crossover. (Left) Relative difference between the numerically minimized hydrodynamic rate function and the conjecture,
\(\delta_f=100\,[f_{\mathrm{num}}(\gamma_0)-f_{\mathrm{conj}}(\gamma_0)]/f_{\mathrm{conj}}(\gamma_0)\).
The numerical result lies approximately \(3\%\)--\(5\%\) above the conjecture throughout the range shown. (Right) Dimensionless critical time \(\mathcal T(\gamma_0)=\hbar\rho_0\tau_c/(mR)\) extracted from the numerical instanton (black squares). The red dashed curve shows the weak-coupling prediction \(\mathcal T(\gamma_0)\simeq2/(\pi\sqrt{\gamma_0})\), while the horizontal dashed line shows the Tonks--Girardeau limit \(\mathcal T\to1/\pi\).}
    \label{fig:crit-time}
\end{figure}

Figure~\ref{fig:conjecture-comparison} shows the numerical results for the dimensionless rate function $f(\gamma_0)$ over more than four orders of magnitude in the interaction strength; details of the numerical method and parameters are given in Appendix~\ref{app:numerical-strat}. The numerical hydrodynamic action follows the conjectured rate function throughout the crossover from weak to strong coupling. We quantify the deviation by the percent difference:
\begin{equation}
    \delta_f(\gamma_0)
    =
    100\times
    \frac{f_{\mathrm{num}}(\gamma_0)-f_{\mathrm{conj}}(\gamma_0)}
    {f_{\mathrm{conj}}(\gamma_0)}.
    \label{eq:delta_f}
\end{equation}
The resulting $\delta_f$ is shown in the left panel of Fig.~\ref{fig:crit-time}. Over the full range of couplings studied, the numerical action lies approximately $3$--$5\%$ above the conjecture, with the deviation remaining roughly constant across the interaction crossover.

This behavior is also seen in the asymptotic regimes discussed above. At weak coupling, the numerical data approach the expected polytropic behavior,
\begin{equation}
    f(\gamma_0)\propto \sqrt{\gamma_0},
\end{equation}
with the logarithmic slope approaching $1/2$. The numerical values remain slightly above the independent weak-coupling result in Eq.~\eqref{eq:weak-hydro-result-main}. Similarly, at strong coupling, the numerical rate approaches the Tonks--Girardeau value $\pi^2/2$ while remaining slightly above the strong-coupling result in Eq.~\eqref{eq:strong-hydro-result-main}.

These independent weak- and strong-coupling results provide useful benchmarks for assessing the numerical accuracy. In both limits, the conjecture agrees with the analytical hydrodynamic expansions, while the numerical results lie systematically above them by an amount comparable to the discrepancy observed throughout the crossover. This suggests that at least part of the systematic \(3\%\)--\(5\%\) offset is numerical. The numerical results therefore support the conjectured rate function across the full interaction crossover at the few-percent level.

To investigate the origin of the residual discrepancy, we perform two numerical convergence tests. First, at the representative coupling $\gamma_0=20$, we increase the resolution of the computational grid from $200\times200$ to $300\times300$. Extrapolation of the late-stage optimization gives an estimated discrepancy of approximately $3.5\%$ on the original grid and $3.0\%$ on the refined grid. The decrease is consistent with convergence toward the conjectured value as the grid resolution is increased. Details of the extrapolation are given in Appendix~\ref{app:numerical-strat}.

Second, at $\gamma_0=1$, we continue the minimization for an additional $20$ optimization cycles. The discrepancy then decreases from approximately $4\%$ to $2.7\%$. Thus, both grid refinement and continued optimization move the numerical action toward the conjectured value, providing evidence that at least part of the residual offset is numerical.

Two representative emptiness instantons are shown in
Fig.~\ref{fig:instanton-examples}. As expected, the empty region persists for a finite imaginary time and has an astroid-like shape. Across the range of couplings studied, the boundary retains this qualitative form and evolves smoothly between the weak- and strong-coupling limits.

The hydrodynamic solution also allows us to extract the dimensionless critical-time function $\mathcal T(\gamma_0)$, shown in Fig.~\ref{fig:crit-time}. Using the results of Ref.~\cite{abanov2025polytropicEFP}, we compare the numerics to the expected weak- and strong-coupling behavior. The numerical \(\mathcal T(\gamma_0)\) reproduces the expected coupling
dependence but lies systematically above both limiting predictions.

The discrepancy in $\tau_c$ is more pronounced than that in the action. Notice that, in Fig.~\ref{fig:instanton-examples}, the density is strongly peaked near the endpoints of the empty interval, $(x,\tau)=(\pm R,0)$, consistent with the expected divergent behavior of the exact solution. In the numerical solution, while there is still a clear density peak, the maximum density remains finite, likely due to the space-time discretization. This provides a possible explanation for the discrepancy in $\tau_c$: regularization of the singular density peaks reduces the local pressure gradient, causing the system to return to equilibrium more slowly and leading to an artificially larger $\tau_c$. We therefore regard the boundary shape and critical-time results primarily as qualitative checks, while the EFP rate function provides the more robust
quantitative comparison with the conjecture.

%%%%%%%%%%%%%%%%%%%%%%%%%%%%%5
%%%%%%%%%%%%%%%%%%%%%%%%%%%%%%%%
\section{Discussion and conclusions}\label{sec:discussion}

We have proposed a parameter-free integral equation for the leading large-interval EFP exponent of the repulsive Lieb--Liniger gas. The conjecture is motivated by a saddle-point analysis of the exact
dual-field Fredholm-determinant representation and is supported by independent hydrodynamic calculations. It reproduces the leading weak- and strong-coupling limits, as well as the first nontrivial correction in both regimes. Numerical minimization of the Lieb--Liniger hydrodynamic action further supports the conjectured rate function over more than four orders of magnitude in $\gamma_0$, with agreement at the few-percent level. Numerical convergence tests indicate that at least part of the residual discrepancy arises from finite grid resolution and incomplete optimization. A complete microscopic derivation would nevertheless require the uniform double-scaling asymptotic analysis identified in Sec.~\ref{sec:conjecture}.

The form of the integral equation has a direct connection to microscopic scattering data. Its logarithmic kernel is the integral over coupling of the
Lieb--Liniger scattering kernel,
\begin{equation}
    \log\left(1+\frac{c^2}{(\lambda-\mu)^2}\right)
    =\int_0^c dc'\,
    \frac{2c'}{(\lambda-\mu)^2+c'^2}.
\end{equation}
Thus the interaction dependence of the proposed EFP exponent is encoded
directly through the two-body scattering data. This observation may be useful in searching for analogous equations in other integrable models, but it does not by itself establish the conjectured large-distance asymptotics.

It would also be interesting to formulate the EFP saddle directly through variables of 
generalized hydrodynamics, where interacting evolution can be
represented as free streaming in state-dependent coordinates
\cite{castro2016emergent,bertini2016ghdXXZ,doyon2018geometry}.

The structure of the terms below the leading $R^2$ exponent remains an open
problem. Introducing the dimensionless interval radius
\[
\ell=\rho_0R,
\]
we consider the asymptotic ansatz
\begin{equation}
    -\log P(R;\gamma_0)
    =a_2(\gamma_0)\ell^2
    +a_1(\gamma_0)\ell
    +b_0(\gamma_0)\log\ell
    +O(1).
    \label{eq:subleading-EFP-ansatz}
\end{equation}
The leading coefficient is the rate function studied in this work,
\[
    a_2(\gamma_0)=f(\gamma_0).
\]
In the Tonks--Girardeau limit, the exact sine-kernel and isotropic XY-chain asymptotics give
$a_1(\infty)=0$ and $b_0(\infty)=1/4$ \cite{Dyson1976Fredholm,Shiroishi2001EFP}. At weak coupling, rescaling
the Gross--Pitaevskii instanton action separates the leading dispersionless
contribution, which is proportional to $\ell^2$, from the quantum-pressure
term, which carries no explicit positive power of $\ell$. This power
counting suggests that the linear coefficient may vanish throughout the
repulsive Lieb--Liniger regime,
\begin{equation}
    a_1(\gamma_0)\stackrel{?}{=}0.
\end{equation}
This is not a proof: the dispersionless instanton is singular at the
endpoints of the empty interval. A careful treatment of these regions is required to determine whether they
produce a term linear in \(\ell\). Alternatively, this question could be
addressed through a uniform finite-coupling asymptotic analysis of the exact
determinant. The logarithmic coefficient
$b_0(\gamma_0)$ likewise remains unknown away from the Tonks--Girardeau
limit.\footnote{A speculative continuum extrapolation of the critical-XXZ EFP
exponent proposed in Ref.~\cite{korepin2003xxz} gives
\[
    b_0(K)\stackrel{?}{=}
    \frac{1}{12}
    +\frac{1}{3}
    \left(\sqrt{2K}-\frac{1}{\sqrt{2K}}\right)^2,
    \qquad
    K=\frac{\pi\rho_0}{v_s}.
\]
Indeed, the XXZ result
\(\gamma_{\rm XXZ}=1/12+\nu^2/[3(1-\nu)]\), together with
\(K_{\rm XXZ}=1/[2(1-\nu)]\), takes precisely this form. It reproduces the
exact Tonks--Girardeau value \(b_0(1)=1/4\). Although the expression is
suggestive of \(c=1\) boundary-CFT formulas, no derivation of its
applicability to the Lieb--Liniger gas is presently known.}

Regardless of the subleading structure, the integral equation proposed here
provides a unified description of the leading EFP exponent. It is supported
by independent analytical results in both coupling limits and by numerical
hydrodynamic calculations across the repulsive interaction crossover.

%%%%%%%%%%%%%%%%%%%%%%%

%%%%%%%%%%%%%%%%%%%%%%%
\section*{Acknowledgments}
%\label{sec:ack}
%%%%%%%%%%%%%%%%%%%%%%%

The authors are grateful to Dmitri Gangardt for fruitful discussions and a careful reading of the manuscript, and Alex Kamenev for insightful comments and helpful suggestions that improved the presentation of this work.

The authors used OpenAI’s GPT‑5.6 Sol through ChatGPT and Codex for assistance with manuscript editing, organization, and the examination of analytical and numerical calculations. All outputs were reviewed by the authors, who take full responsibility for the manuscript.
%\paragraph{Funding information}
%AGA's work was supported by the National Science Foundation under Grant NSF DMR--2116767.

%%%%%%%%%%%%%%%%%%%%%%%%%%%%%%%%%
%%%%%%%%%%%%%%%%%%%%%%%%%%%%%%%%%
%\bibliographystyle{unsrt}

%%%%%%%%%%%%%%%%%%%%%%%%%%%%%%%%%%%%%

\section*{Data and code availability}
The code and data required to reproduce the main numerical results and figures of this work are openly available at
\url{https://github.com/bkhanikati/EFP_lieb_liniger.git}.
The repository contains the code used to compute the hydrodynamic action and the resulting numerical data. The additional convergence tests presented in the text and in Appendix~\ref{app:numerical-strat} were performed using the same numerical procedure, with the grid resolutions and numbers of optimization cycles specified therein.

%%%%%%%%%%%%%%%%%%%%%%%%%%%%%%%%%%%%%%%

\newpage
\bibliography{emptiness}

%\appendix

%%%%%%%%%%%%%%%%%%%%%%%
\begin{appendix}
%%%%%%%%%%%%%%%%%%%%%%%

%%%%%%%%%%%%%%%%%%%%%%%
\section{Numerical strategy}\label{app:numerical-strat}
Here, we present the numerical method used to calculate the emptiness
instanton. We minimize the background-subtracted Euclidean action introduced
in Eq.~\eqref{eq:hydro-action},
\begin{equation}
    S_E = \int dx\,d\tau \left[
        \frac{1}{2}\rho v^2
        + \epsilon(\rho)
        - \epsilon(\rho_0)
        - \mu_0(\rho-\rho_0)
    \right].
\end{equation}
Defining
\begin{equation}
    h(\rho)
    =
    \epsilon(\rho)-\epsilon(\rho_0)-\mu_0(\rho-\rho_0),
\end{equation}
the uniform far-field state contributes zero to the action.

\subsection{Equation of state}

The energy density is obtained from the Lieb--Liniger ground state,
\begin{equation}
    \epsilon(\rho)
    =
    \int_{-q}^{q} d\theta\,
    \frac{\theta^2}{2}\rho_p(\theta),
\end{equation}
where $q$ is fixed by the density normalization. For each value of the
coupling used in the hydrodynamic calculation, we construct the equation of
state numerically by solving the Lieb equation \eqref{eq:Lieb-equation} using product integration of
the Lorentzian kernel, with the kernel integrated analytically over each
rapidity panel. The calculations use $N_q=900$ points
for the rapidity cutoff and $N_\theta=1400$ rapidity panels. The resulting
equation-of-state table extends to densities $20\rho_0$ and is interpolated
during the hydrodynamic minimization. The resulting equation of state is
illustrated for $c=1$ in Fig.~\ref{fig:LL-eos}, together with its limiting
weak- and strong-coupling density dependences.

\subsection{Boundary-fitted formulation of the action}

We optimize both the boundary shape and the hydrodynamic field. The initial
guess $\tau_c^{(0)}$ for the critical time $\tau_c$ is chosen using the sound
velocity,
\begin{equation}
    v_s^2 = \rho_0\epsilon''(\rho_0),
    \qquad
    \tau_c^{(0)} = \frac{R}{v_s}.
\end{equation}

It is sufficient to optimize the action in the first quadrant due to parity
and time-reversal symmetry. We write
\begin{equation}
    x = X(\zeta)+y,
    \qquad
    \tau = \tau_c \zeta^3,
\end{equation}
where $y\geq0$. The curve $y=0$ with $0\leq \zeta\leq1$ defines the boundary
of the instanton. Thus $\zeta=0$, $\tau=0$, and $X(0)=R$ define the bottom of
the instanton boundary, while $\zeta=1$, $\tau=\tau_c$, and $X(1)=0$ define
its top.

For the numerical results presented in the main text, we set
$R=1$ and $\rho_0=1$. The exterior hydrodynamic domain is discretized on
a $200\times200$ grid in the boundary-fitted coordinates $(\zeta,y)$. The
transverse coordinate extends to
$
    y_{\max}=7R,
$
while the imaginary-time domain extends to
$
    \tau_{\max}=7\tau_c.
$
Equivalently, the computational coordinate extends to
$\zeta_{\max}=7^{1/3}$. Approximately $72\%$ of the $\zeta$-grid points are
placed in the interval $0\leq \zeta\leq1$, thereby concentrating the temporal resolution in the
region containing the boundary of the empty domain.

Introducing again the displacement field $u(x,\tau)$, with
\begin{equation}
    \rho=u_x,
    \qquad
    \rho v=-u_\tau,
\end{equation}
we define
\begin{equation}
    U(y,\zeta)
    =
    u\bigl(X(\zeta)+y,\tau_c \zeta^3\bigr).
\end{equation}
Then $\rho=U_y$. By the chain rule,
\begin{equation}
    u_\tau
    =
    \frac{U_\zeta-X_\zeta\rho}{\tau_\zeta}.
\end{equation}
Defining
\begin{equation}
    A=U_\zeta-X_\zeta\rho,
\end{equation}
we have
\begin{equation}
    u_\tau=\frac{A}{\tau_\zeta}.
\end{equation}
The bulk contribution to the action becomes
\begin{equation}
    S_b
    =
    \int d\zeta\,dy
    \left[
        \frac{A^2}{2\tau_\zeta\rho}
        +\tau_\zeta h(\rho)
    \right].
\end{equation}

Since the action is written relative to the uniform background, the excluded
empty region also contributes a constant background term. Namely,
\begin{equation}
    h(0)=\mu_0\rho_0-\epsilon(\rho_0).
\end{equation}
The contribution from the empty region is therefore
\begin{equation}
    S_h
    =
    h(0)\int_0^1 d\zeta\,\tau_\zeta X(\zeta).
\end{equation}
The Euclidean action in one quadrant is
\begin{equation}
    S_E^{(1)}=S_b+S_h.
\end{equation}
The full action is then
\begin{equation}
    S_E=4S_E^{(1)},
\end{equation}
where the factor of $4$ accounts for the four quadrants. In the units used
for the numerical calculations, $R=\rho_0=1$, the resulting full action is
equal to the dimensionless rate function $f(\gamma_0)$ quoted in the main
text.

The $y$ coordinate is represented by cell midpoints on a uniform grid,
whereas the nonuniform $\zeta$ grid is represented by nodes and adjacent
cell midpoints. The displacement $U$ is reconstructed by cumulative midpoint
quadrature in $y$. Derivatives with respect to $\zeta$ are evaluated by
finite differences between neighboring nodes, and the transformed action is
integrated by midpoint quadrature over the resulting cells.

\subsection{Variational parametrization and minimization}

To ensure positivity of the density, we parameterize it as
\begin{equation}
    \rho(\zeta,y)
    =
    \rho_0 B(\zeta,y)
    \exp\left[E(y)\phi(\zeta,y)\right].
\end{equation}
Here, $\phi(\zeta,y)$ is the main field being optimized. The function $B(\zeta,y)$ is a fixed depletion profile,
\begin{equation}
    B(\zeta,y)
    =
    1-H(\zeta)\exp\left[-\left(\frac{y}{\ell_h}\right)^2\right]
\end{equation}
with $\ell_h=0.12R$, where
\begin{equation}
    H(\zeta)
    =
    \frac{1}{1+\exp[(\zeta-1)/\Delta \zeta]},
    \qquad
    \Delta \zeta=\frac{3}{N_{\rm hole}},
\end{equation}
and $N_{\rm hole}=0.72N_\zeta$. The function $H(\zeta)$ smoothly switches off
the depletion profile above the empty region. The envelope is chosen as
\begin{equation}
    E(y)
    =
    \left(1-\frac{y}{y_{\max}}\right)^2.
\end{equation}
Thus, $B(\zeta,y)$ provides the short-distance depletion near the emptiness
boundary, while $E(y)$ forces the optimized field to vanish at the outer
edge of the numerical domain, so that $\rho\to\rho_0$ as
$y\to y_{\max}$.
Once the density is known, the displacement
field is reconstructed by integrating
\begin{equation}
    U(\zeta,y)
    =
    \int_0^y \rho(\zeta,y')\,dy'.
\end{equation}

The displacement field must also reproduce the density removed from the
empty interval. On each constant-$\zeta$ slice this gives the constraint
\begin{equation}
    \int_0^{y_{\max}}
    \left[\rho(\zeta,y)-\rho_0\right]dy
    =\rho_0X(\zeta),
    \label{eq:numerical-displacement-constraint}
\end{equation}
up to corrections associated with truncating the exterior domain at
$y=y_{\max}$. This condition is imposed through the numerical objective
described below.

The boundary shape is optimized together with the hydrodynamic field. We
introduce a positive boundary speed through
\begin{equation}
    X(\zeta)
    =
    R\left[1-\frac{I(\zeta)}{I(1)}\right],
    \qquad
    I(\zeta)=\int_0^\zeta w(s)\,ds,
\end{equation}
with $w(s)>0$. This automatically enforces
$X(0)=R$, $X(1)=0$, and $X_\zeta\leq0$. As an initial guess, we use the
free-fermion boundary
\begin{equation}
    X_0(\zeta)=R(1-\zeta^2)^{3/2},
\end{equation}
and write the optimized boundary speed as
\begin{equation}
    w(s)=w_0(s)e^{\eta(s)}.
\end{equation}
The deformation is expanded in $M=10$ Chebyshev modes,
\begin{equation}
    \eta(s)
    =
    \eta_{\max}
    \tanh\left[
        \frac{1}{\eta_{\max}}
        \sum_{m=1}^{10}a_m T_m(2s-1)
    \right],
\end{equation}
where $\eta_{\max}=1$ in the numerical calculations.

The critical time $\tau_c$ is optimized together with the boundary and the
hydrodynamic field. We introduce
\begin{equation}
    \ell_\tau=\log\tau_c,
\end{equation}
which automatically enforces $\tau_c>0$.

The numerical minimization is performed in double precision. We first use the Adam optimizer \cite{kingma2015adam} to relax the density field, boundary coefficients, and critical time from the initial configuration. The resulting configuration is subsequently refined using the L-BFGS algorithm with a strong-Wolfe line search.

The numerical objective has the form
\begin{equation}
    \mathcal L
    =S_E^{(1)}+\sum_j W_j\mathcal P_j,
    \label{eq:numerical-objective}
\end{equation}
where the $\mathcal P_j$ are non-negative penalty and regularization terms.
The principal penalty enforces
Eq.~\eqref{eq:numerical-displacement-constraint}. Further penalties enforce
relaxation to $\rho=\rho_0$ and vanishing current at the outer boundaries,
time-reflection symmetry at $\tau=0$, and vanishing normal derivatives and
velocity on the symmetry axis above the instanton. Weak regularization terms
control short-wavelength variations of the density field and boundary shape.
A soft overflow penalty discourages the density from leaving the range covered
by the equation-of-state table. The action quoted in the main text is the
physical full action $S_E$, evaluated on the optimized configuration, rather
than the value of the auxiliary objective $\mathcal L$.

The optimization is performed by continuation. The initial density profile
is constructed to approximately satisfy the displacement constraint, with
the free-fermion boundary and $\tau_c^{(0)}=R/v_s$ used as seeds. Alternating
Adam and L-BFGS stages then optimize the density field, the Chebyshev
coefficients of the boundary, and $\log\tau_c$ simultaneously. The best
configuration reached during each stage is used to initialize the next one.

As diagnostics, we monitor the physical action, the penalty contribution,
the displacement-constraint residual, the far-boundary and symmetry
residuals, and the ratio of the kinetic and potential contributions to the
action.

%%%%%%%%%
\subsection{Convergence tests and numerical accuracy}

We perform additional numerical tests to assess whether the systematic residual discrepancy is attributable to finite grid resolution and incomplete optimization convergence.

\begin{figure}[H]
    \centering
    \includegraphics[width=1\linewidth]{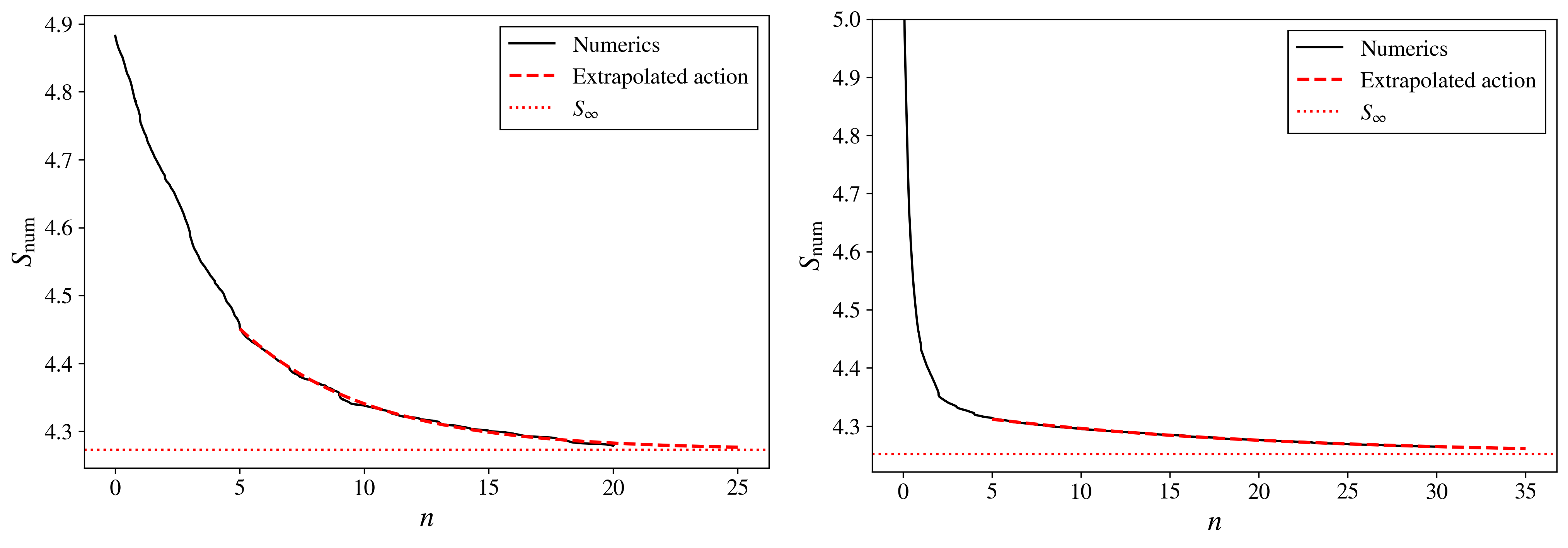}
    \caption{Late-stage optimization history of the physical action at
    $\gamma_0=20$ for the $200\times200$ and $300\times300$ computational
    grids. The black curve shows the numerical action, while the dashed curves show fits to
    $S(n)=S_\infty+A e^{-n/\tau}$. For the $200 \times 200$ grid, the fitting window is over $n =$ 5--20; for $300\times 300$, it is $n =$5--30.}
    \label{fig:optimization-history}
\end{figure}

We first test convergence with respect to the computational grid resolution at the representative point $\gamma_0=20$. We compare calculations performed on the original $200\times200$ grid with a grid refined by $50\%$ along each axis, corresponding to $300\times300$ grid points. To account for the possibility that the optimization is not fully converged within the available number of iterations, we estimate the late-iteration asymptotic value of the action by fitting the final portion of the optimization history to the empirical form
\begin{equation}
    S(n) = S_{\infty} + A \exp\left(-\frac{n}{\tau}\right),
\end{equation}
where $S_{\infty}$ is the extrapolated asymptotic value, $A$ and $\tau$ describe the remaining optimization error, and $n$ is the number of L-BFGS optimization blocks (with $10^3$ iterations per block). 
The optimization histories and the corresponding empirical fits are shown
in Fig.~\ref{fig:optimization-history}. 

We exclude the transient early stages and restrict the fit to the region in which the decrease of the action is approximately exponential. We also vary the number of late optimization cycles included in the fit to assess the sensitivity of the extrapolation to the fitting interval. For the $200\times200$ calculation, the late-stage fit gives
\begin{equation}
    S_{\infty}^{200\times200} \simeq 4.273,
\end{equation}
while for the refined $300\times300$ calculation we obtain
\begin{equation}
    S_{\infty}^{300\times300} \simeq 4.251.
\end{equation}
For comparison, the conjectured value at $\gamma_0=20$ is
$S\simeq 4.129$. The corresponding discrepancies are approximately $3.5\%$
and $3.0\%$, respectively. The decrease under grid refinement is consistent
with a finite-resolution contribution to the residual numerical error. These
are coarse extrapolated estimates rather than continuum-extrapolated results. The directly computed action on the $300 \times 300$ grid is also smaller than on the $200 \times 200$ grid, with $S \approx 4.265$ and $S \approx 4.279$, respectively. 

As a complementary test, we examine convergence under continued
optimization. At $\gamma_0=1$, performing an additional $20$ optimization
cycles, with up to $5000$ L-BFGS iterations per cycle, reduces the relative
discrepancy from approximately $4\%$ to approximately $2.7\%$. This continued
decrease indicates that incomplete optimization convergence contributes to
the numerical discrepancy.

Taken together, the grid-refinement and continued-optimization tests show
systematic improvement at different interaction strengths as the numerical calculation becomes better resolved.
Although these tests do not establish the fully converged continuum value,
they indicate that part of the remaining discrepancy can be attributed to numerical limitations.

%%%%%%%%%%
\section{First-order expansions}\label{app:perturb}

%%%%%%%%%%
\subsection{Strong-coupling expansion}
We derive the first-order correction to the strong-coupling EFP by mapping
the Euler-scale hydrodynamic action to its free-fermion form. The mapping is the signed-length analogue of the excluded-volume transformation for hard
rods and is closely related to the geometric formulation of generalized hydrodynamics \cite{doyon2018geometry}.

For $c\gg\rho$, the Lieb--Liniger scattering kernel has the expansion
\begin{equation}
    \Delta(\theta)
    =\frac{2}{c}-\frac{2\theta^2}{c^3}+\cdots.
\end{equation}
Keeping the constant term in the Lieb equation \eqref{eq:Lieb-equation} gives
\begin{equation}
    \rho_p(\theta)
    =\frac{1}{2\pi}
    \left(1+\frac{2\rho}{c}\right)+\cdots.
    \label{eq:strong-rhop}
\end{equation}
The density normalization and the ground-state energy density are therefore
\begin{align}
    q
    &=\frac{\pi\rho}{1+2\rho/c}
      =\pi\rho\left(1-\frac{2\rho}{c}\right)
      +O\!\left(\frac{\rho^3}{c^2}\right),
      \label{eq:strong-q}\\
    \epsilon(\rho)
    &=\frac{\pi^2\rho^3}{6}
      \left(1-\frac{4\rho}{c}\right)
      +O\!\left(\frac{\rho^5}{c^2}\right).
      \label{eq:strong-eos}
\end{align}

We now introduce an expanded-volume coordinate $y$ and a transformed density
$n(y,\tau)$. At fixed $\tau$, the change of coordinates is defined by
\begin{equation}
    x(y,\tau)
    =y-\frac{2}{c}\int^{y}n(y',\tau)\,dy',
    \qquad
    \frac{\partial x}{\partial y}=1-\frac{2n}{c}.
    \label{eq:strong-coordinate-map}
\end{equation}
The additive constant in $x$ is fixed by anchoring the two coordinates in the
empty region. Conservation of particle number gives
\begin{equation}
    \rho(x,\tau)\,dx=n(y,\tau)\,dy,
\end{equation}
and hence
\begin{equation}
    \rho=\frac{n}{1-2n/c}
    =n+\frac{2n^2}{c}+O\!\left(\frac{n^3}{c^2}\right),
    \qquad
    n=\rho-\frac{2\rho^2}{c}+O\!\left(\frac{\rho^3}{c^2}\right).
    \label{eq:strong-density-map}
\end{equation}
Thus Eq.~\eqref{eq:strong-coordinate-map} corresponds to a signed effective
rod length $a_{\rm eff}=-2/c$.

The transformation also applies to time-dependent configurations. The
integral in Eq.~\eqref{eq:strong-coordinate-map} is the cumulative particle
number, which is constant along a fluid trajectory. The velocity of a fluid
element is consequently the same in the $x$ and $y$ coordinates. The kinetic
part of the Euclidean action then obeys
\begin{equation}
    dx\,\frac{\rho v^2}{2}
    =dy\,\frac{n v^2}{2}.
    \label{eq:strong-kinetic-map}
\end{equation}
Using Eqs.~\eqref{eq:strong-eos} and \eqref{eq:strong-density-map}, the energy
term becomes
\begin{align}
    dx\,\epsilon(\rho)
    &=dy\left(1-\frac{2n}{c}\right)
    \frac{\pi^2}{6}
    \left(\frac{n}{1-2n/c}\right)^3
    \left(1-\frac{4n}{c(1-2n/c)}\right)
    +O\!\left(\frac{n^5}{c^2}\right)dy\nonumber\\
    &=dy\,\frac{\pi^2n^3}{6}
      +O\!\left(\frac{n^5}{c^2}\right)dy.
    \label{eq:strong-energy-map}
\end{align}
For a finite system at fixed particle number, the uniform background action
transforms in the same way; taking the thermodynamic limit therefore maps the
action to the free-fermion one through first order in $1/c$.

The transformed uniform density is
\begin{equation}
    n_0
    =\rho_0-\frac{2\rho_0^2}{c}
    +O\!\left(\frac{\rho_0^3}{c^2}\right)
    =\rho_0\left(1-\frac{2}{\gamma_0}
    +O(\gamma_0^{-2})\right).
    \label{eq:strong-background-density}
\end{equation}
Inside the empty interval $n=0$, so the coordinate map has unit Jacobian and
its radius remains $R$, up to an $R$-independent shift that cannot affect the
leading $R^2$ exponent. The free-fermion result in the transformed variables
therefore gives
\begin{align}
    -\log P(R)
    &=\frac{\pi^2}{2}(n_0R)^2+o(R^2)\nonumber\\
    &=\frac{\pi^2}{2}(\rho_0R)^2
      \left[1-\frac{4}{\gamma_0}
      +O(\gamma_0^{-2})\right]+o(R^2).
\end{align}
Thus the strong-coupling rate function is
\begin{equation}
    f(\gamma_0)
    =\frac{\pi^2}{2}
    \left[1-\frac{4}{\gamma_0}
    +O(\gamma_0^{-2})\right].
    \label{eq:strong-hydro-appendix}
\end{equation}

%%%%%%%%
\subsection{Weak-coupling expansion}
At weak coupling, the first correction can be found by first-order
perturbation theory around the exactly known polytropic instanton with
$\gamma_{\rm pol}=2$.  The Lieb--Liniger equation of state has the expansion
\begin{equation}
    \epsilon(\rho)
    =
    \frac{c\rho^2}{2}
    -\frac{2}{3\pi}c^{3/2}\rho^{3/2}
    +O(c^2\rho).
\end{equation}
We introduce dimensionless variables
\begin{equation}
    \rho=\rho_0 \varrho,
    \qquad
    x=RX,
    \qquad
    \tau=\frac{R}{\rho_0\sqrt{\gamma_0}}T,
    \qquad
    v=\rho_0\sqrt{\gamma_0}\,u,
\end{equation}
where $\gamma_0=c/\rho_0$. In these variables, the continuity equation and the EFP boundary conditions are
\begin{equation}
    \partial_T \varrho+\partial_X(\varrho u)=0,
    \qquad
    \varrho(X,0)=0\quad (|X|<1),
    \qquad
    \varrho\to1,\quad u\to0
    \quad\text{as }X^2+T^2\to\infty.
\end{equation}
We denote by $\mathcal A$ the set of fields satisfying these constraints.

To make the background subtraction explicit, define
\begin{equation}
    h(\rho)=\epsilon(\rho)-\epsilon(\rho_0)
    -\mu_0(\rho-\rho_0),
    \qquad
    \mu_0=\epsilon'(\rho_0).
\end{equation}
Using the weak-coupling equation of state, we find
\begin{equation}
    \frac{h(\rho_0\varrho)}{\rho_0^3}
    =
    \frac{\gamma_0}{2}(\varrho-1)^2
    +\gamma_0^{3/2}H_1(\varrho)
    +O(\gamma_0^2),
\end{equation}
where
\begin{equation}
    H_1(\varrho)
    =
    \frac{1}{\pi}
    \left(
        -\frac{2}{3}\varrho^{3/2}+\varrho-\frac{1}{3}
    \right).
\end{equation}
The spacetime measure transforms as
\begin{equation}
    dx\,d\tau
    =\frac{R^2}{\rho_0\sqrt{\gamma_0}}\,dX\,dT.
\end{equation}
After dividing the Euclidean action by $(\rho_0R)^2$, the dimensionless EFP
rate functional takes the form
\begin{equation}
    f(\gamma_0)
    =
    \min_{(\varrho,u)\in\mathcal A}
    \left\{
        \sqrt{\gamma_0}\,\mathcal I_0[\varrho,u]
        +\gamma_0\,\mathcal I_1[\varrho]
        +O(\gamma_0^{3/2})
    \right\},
\end{equation}
with
\begin{align}
    \mathcal I_0[\varrho,u]
    &=
    \int dX\,dT
    \left[
        \frac{1}{2}\varrho u^2+\frac{1}{2}(\varrho-1)^2
    \right],\\
    \mathcal I_1[\varrho]
    &=
    \int dX\,dT\,H_1(\varrho),
\end{align}
where the first functional contains the mean-field action and the second is
the explicit Lee--Huang--Yang correction.

Let $(\varrho_0,u_0)$ denote the exact $\gamma_{\rm pol}=2$ emptiness instanton.
To see why no correction to the fields is required at this order, set
$s=\sqrt{\gamma_0}$ and write the perturbed saddle as
\begin{equation}
    (\varrho_s,u_s)=(\varrho_0,u_0)+s(\varrho_1,u_1)+O(s^2).
\end{equation}
Expanding the on-shell functional gives
\begin{align}
    f(\gamma_0)
    ={}&s\mathcal I_0[\varrho_0,u_0]
    +s^2\left\{
        \mathcal I_1[\varrho_0]
        +\delta\mathcal I_0[\varrho_0,u_0]\cdot(\varrho_1,u_1)
    \right\}
    +O(s^3).
\end{align}
The last term in braces vanishes because $(\varrho_0,u_0)$ is stationary under
variations in $\mathcal A$.  This includes variations of the vacuum boundary:
the leading instanton satisfies both the bulk Euler equations and the
free-boundary transversality condition.  Consequently,
\begin{equation}
    f(\gamma_0)
    =
    \frac{16}{3\pi}\sqrt{\gamma_0}
    +A_1\gamma_0
    +O(\gamma_0^{3/2}),
\end{equation}
where
\begin{equation}
    A_1
    =
    \int_{\mathbb R^2}dX\,dT\,
    H_1\!\left(\varrho_0(X,T)\right).
    \label{eq:weak-A1}
\end{equation}
Before performing the analytical reduction below, direct numerical quadrature
of Eq.~\eqref{eq:weak-A1} on the exact polytropic instanton gives
\begin{equation}
    A_1^{\rm num}\simeq-0.15656.
    \label{eq:weak-A1-numerical}
\end{equation}

The local weak-coupling expansion is formally nonuniform at the vacuum
boundary because the local interaction parameter is
$c/\rho=\gamma_0/\varrho$. This does not modify $A_1$. The constant and linear
parts of the background-subtracted energy are fixed by $\epsilon(\rho_0)$
and $\mu_0$ and are already included in $H_1$. In particular,
\begin{equation}
    H_1(0)=-\frac{1}{3\pi}.
\end{equation}
In the crossover region $\varrho=O(\gamma_0)$, the nonlinear part of both the
exact energy density and its weak-coupling approximation is
$O(\rho_0^3\gamma_0^3)$.  Even if the corresponding dimensionless spacetime
area is only bounded by $O(1)$, the factor $\gamma_0^{-1/2}$ in the
spacetime measure places a possible mismatch at $O(\gamma_0^{5/2})$, well
beyond the order considered here.

We now evaluate Eq.~\eqref{eq:weak-A1} on the analytic
$\gamma_{\rm pol}=2$ instanton \cite{abanov2025polytropicEFP}.  In the
conventions of that solution, $t=2T$, and the Riemann invariants are
\begin{equation}
    \lambda=a+ib,
    \qquad
    \bar{\lambda}=a-ib,
    \qquad
    \varrho=b^2,
    \qquad
    u=2a.
\end{equation}
Writing
\begin{equation}
    F(a,b)=\partial_\lambda V_{1/2}(\lambda,\bar{\lambda}),
\end{equation}
where $\bar\lambda$ is held fixed in the derivative, the hodograph map on
the physical branch is
\begin{equation}
    t=-\frac{2}{b}\operatorname{Im}F,
    \qquad
    X=\operatorname{Re}F+at.
\end{equation}
The physical sheet is chosen so that $a\in\mathbb R$ and $b>0$ cover the
fluid region in the lower Euclidean-time half-plane once.  Its Jacobian is
\begin{equation}
    J(a,b)
    =
    \det\frac{\partial(X,t)}{\partial(a,b)}
    =
    X_a t_b-X_b t_a.
\end{equation}
Since $t=2T$,
\begin{equation}
    dX\,dT=\frac{1}{2}|J(a,b)|\,da\,db.
\end{equation}
Reflection to the upper half-plane cancels the factor $1/2$.  The fluid
contribution is therefore
\begin{equation}
    A_1^{\rm fluid}
    =
    \int_{-\infty}^{\infty}da
    \int_0^\infty db\,
    H_1(b^2)\,|J(a,b)|.
\end{equation}
Using
\begin{equation}
    H_1(b^2)
    =
    -\frac{(b-1)^2(2b+1)}{3\pi},
\end{equation}
we obtain
\begin{equation}
    A_1^{\rm fluid}
    =
    -\frac{1}{3\pi}
    \int_{-\infty}^{\infty}da
    \int_0^\infty db\,
    (b-1)^2(2b+1)|J(a,b)|.
\end{equation}
Its analytic evaluation is most conveniently performed together with the
vacuum contribution, as shown in Appendix~\ref{app:weak-integral}.

The hodograph coordinates cover only the fluid region, so the empty region
must be included separately. Since
\begin{equation}
    H_1(0)=-\frac{1}{3\pi},
\end{equation}
its contribution is fixed by the spacetime area of the vacuum.  For the
$\gamma_{\rm pol}=2$ instanton, the right boundary in the lower-time
half-plane is
\begin{equation}
    t=-\chi'(v),
    \qquad
    X=\chi(v)-v\chi'(v),
\end{equation}
with
\begin{equation}
    \chi'(v)=\frac{4}{\pi(1+v^2)^2},
    \qquad
    \chi(v)=\frac{2}{\pi}
    \left(
        \arctan v+\frac{v}{1+v^2}
    \right).
\end{equation}
Thus $dt=-\chi''(v)\,dv$. Reflecting the right, lower quadrant in space and
time gives
\begin{equation}
    \mathcal A_{\rm vac}^{(t)}
    =
    4\int_0^\infty
    [\chi(v)-v\chi'(v)][-\chi''(v)]\,dv
    =
    \frac{5}{\pi}.
\end{equation}
Since $T=t/2$, the area in the convention of Eq.~\eqref{eq:weak-A1} is
\begin{equation}
    \mathcal A_{\rm vac}^{(T)}
    =
    \frac{5}{2\pi},
\end{equation}
and therefore
\begin{equation}
    A_1^{\rm vac}
    =
    H_1(0)\mathcal A_{\rm vac}^{(T)}
    =
    -\frac{5}{6\pi^2}.
\end{equation}

The exact integral calculation in Appendix~\ref{app:weak-integral} gives
\begin{equation}
    A_1^{\rm fluid}
    =\frac{29/6-8\log 2}{\pi^2},
\end{equation}
and hence
\begin{equation}
    A_1=A_1^{\rm fluid}+A_1^{\rm vac}
    =\frac{4(1-\log 4)}{\pi^2}
    .
    \label{eq:weak-A1-exact}
\end{equation}
The exact value is $A_1=-0.15655920\ldots$, in agreement with the direct
numerical evaluation in Eq.~\eqref{eq:weak-A1-numerical}. The weak-coupling rate obtained
directly from the polytropic instanton is therefore
\begin{equation}
    f(\gamma_0)
    =
    \frac{16}{3\pi}\sqrt{\gamma_0}
    +\frac{4(1-\log 4)}{\pi^2}\gamma_0
    +O(\gamma_0^{3/2}).
\end{equation}

%%%%%%%%%%%%
\section{Integral computation for the weak-coupling correction}
\label{app:weak-integral}

Here we evaluate Eq.~\eqref{eq:weak-A1} on the
$\gamma_{\rm pol}=2$ instanton. We use the notation
\begin{equation}
    F=P+iQ,
    \qquad
    h(b)=H_1(b^2)
    =-\frac{(b-1)^2(2b+1)}{3\pi}.
\end{equation}
The hodograph map is
\begin{equation}
    t=-\frac{2Q}{b},
    \qquad
    X=P+at,
\end{equation}
and its Jacobian,
\begin{equation}
    J=X_a t_b-X_b t_a,
\end{equation}
is negative on the physical sheet. Consequently,
\begin{equation}
    A_1^{\rm fluid}
    =
    -\int_{-\infty}^{\infty}da
    \int_0^\infty db\,h(b)J(a,b).
    \label{eq:A1-fluid-oriented}
\end{equation}

We first reduce this expression by integration by parts. Since \(h\) depends
only on \(b\),
\begin{equation}
    hJ
    =
    \partial_a\!\left(hXt_b\right)
    -\partial_b\!\left(hXt_a\right)
    +h'Xt_a.
    \label{eq:A1-total-derivative}
\end{equation}
Moreover,
\begin{equation}
    h'(b)=\frac{2b(1-b)}{\pi},
    \qquad
    t_a=-\frac{2}{b}\operatorname{Im}F_a,
\end{equation}
and therefore the bulk term in Eq.~\eqref{eq:A1-fluid-oriented} becomes
\begin{equation}
    -h'Xt_a
    =
    \frac{4}{\pi}(1-b)X\,\operatorname{Im}F_a.
    \label{eq:A1-reduced-bulk}
\end{equation}

For completeness, we now justify the boundary terms generated by
Eq.~\eqref{eq:A1-total-derivative}. We perform the integration on the
regulated domain
\begin{equation}
    \Omega_{A,B,\varepsilon}
    =
    \left\{
        (a,b): |a|<A,\ 0<b<B
    \right\}
    \setminus D_\varepsilon(0,1),
    \label{eq:A1-regulated-domain}
\end{equation}
where \(D_\varepsilon(0,1)\) is a small disk around the point
\((a,b)=(0,1)\), which represents spacetime infinity. We estimate the
vertical and horizontal parts of the regulated boundary separately before
taking $A,B\to\infty$, and then take $\varepsilon\to0$.

At \(b=0\), the boundary term is
\begin{equation}
    \mathcal C_0
    =
    -\int_{-\infty}^{\infty}
    h(0)X(a,0)t_a(a,0)\,da
    =
    -h(0)\oint_{\partial\mathcal V}X\,dt,
    \label{eq:A1-b0-boundary}
\end{equation}
where \(\partial\mathcal V\) is the boundary of the empty region, oriented
counterclockwise. Since
\begin{equation}
    \oint_{\partial\mathcal V}X\,dt
    =
    \mathcal A_{\rm vac}^{(T)}
    =
    \frac{5}{2\pi},
\end{equation}
and \(h(0)=-1/(3\pi)\), one obtains
\begin{equation}
    \mathcal C_0
    =
    \frac{5}{6\pi^2}
    =
    -A_1^{\rm vac}.
    \label{eq:A1-vacuum-cancellation}
\end{equation}
Thus the \(b=0\) contribution generated by integration by parts cancels
exactly the separately included vacuum contribution.

The remaining outer boundaries vanish. Direct expansion of the branchwise
integral representations of the hodograph solution given in
Ref.~\cite{abanov2025polytropicEFP} gives, for large \(\lvert a\rvert\) at
fixed \(b\),
\begin{equation}
    X(a,b)
    =
    \operatorname{sgn}(a)+O(|a|^{-3}),
    \qquad
    t(a,b)=O(|a|^{-4}),
    \qquad
    t_b(a,b)=O(|a|^{-5}).
    \label{eq:A1-large-a}
\end{equation}
Hence, for fixed \(B\),
\begin{equation}
    \int_0^B h(b)X(\pm A,b)t_b(\pm A,b)\,db
    =
    O(A^{-5})
    \longrightarrow0.
    \label{eq:A1-large-a-boundary}
\end{equation}

On the high-density branch, expansion of the corresponding representation in
Ref.~\cite{abanov2025polytropicEFP} gives, uniformly for \(a\) in a bounded
interval,
\begin{equation}
    F(a,b)
    =
    1+O(b^{-3})+i\,O(ab^{-4}).
    \label{eq:A1-large-b-F}
\end{equation}
It follows that
\begin{equation}
    X(a,b)=1+O(b^{-3}),
    \qquad
    t_a(a,b)=O(b^{-5}).
\end{equation}
Although \(h(b)=-2b^3/(3\pi)+O(b^2)\), their product satisfies
\begin{equation}
    h(b)X(a,b)t_a(a,b)=O(b^{-2}),
\end{equation}
so that, for fixed \(A\),
\begin{equation}
    \int_{-A}^{A}h(B)X(a,B)t_a(a,B)\,da
    =
    O(B^{-2})
    \longrightarrow0.
    \label{eq:A1-large-b-boundary}
\end{equation}

Finally, consider the small contour surrounding \((a,b)=(0,1)\). Writing
\begin{equation}
    \delta=\sqrt{a^2+(b-1)^2},
\end{equation}
the far-field expansion of the hodograph solution gives
\begin{equation}
    X,t=O(\delta^{-1/2}),
    \qquad
    \nabla_{a,b}t=O(\delta^{-3/2}).
    \label{eq:A1-matching-asymptotics}
\end{equation}
At the same time,
\begin{equation}
    h(b)
    =
    -\frac{(b-1)^2}{\pi}
    +O\!\left((b-1)^3\right)
    =
    O(\delta^2).
\end{equation}
The boundary current in Eq.~\eqref{eq:A1-total-derivative} is therefore
\(O(1)\), while the length of \(\partial D_\varepsilon(0,1)\) is
\(O(\varepsilon)\). With the clockwise orientation induced on this inner
boundary, its contribution consequently satisfies
\begin{equation}
    \oint_{\partial D_\varepsilon(0,1)}
    \left(hXt_b\,db+hXt_a\,da\right)
    =
    O(\varepsilon)
    \longrightarrow0.
    \label{eq:A1-matching-boundary}
\end{equation}

Combining Eqs.~\eqref{eq:A1-reduced-bulk}--\eqref{eq:A1-matching-boundary},
and including the vacuum contribution, gives
\begin{equation}
    A_1
    =
    \frac{4}{\pi}
    \int_{-\infty}^{\infty}da
    \int_0^\infty db\,
    (1-b)X\,\operatorname{Im}F_a.
    \label{eq:A1-before-parseval}
\end{equation}

The hodograph function \(F\) for the polytropic instanton with \(\gamma_{\rm pol}=2\) satisfies the differential relation \cite{abanov2025polytropicEFP}
\begin{equation}
    F_{\bar\lambda}
    =
    \frac{F-\bar F}{2(\lambda-\bar\lambda)}.
    \label{eq:F-differential-relation}
\end{equation}
Writing \(F=P+iQ\) and \(\lambda=a+ib\), this is equivalently
\begin{equation}
    P_a-Q_b=\frac{Q}{b},
    \qquad
    Q_a+P_b=0.
    \label{eq:F-PQ-relations}
\end{equation}
To see explicitly how the Bessel functions arise, we use the Fourier
convention
\begin{equation}
    \widehat P(k,b)=\int_{-\infty}^{\infty}e^{-ika}P(a,b)\,da,
    \qquad
    \widehat Q(k,b)=\int_{-\infty}^{\infty}e^{-ika}Q(a,b)\,da.
\end{equation}
For every nonzero Fourier mode, Eqs.~\eqref{eq:F-PQ-relations} imply
\begin{align}
    \widehat P_{bb}+\frac{1}{b}\widehat P_b-k^2\widehat P&=0,
    \\
    \widehat Q_{bb}+\frac{1}{b}\widehat Q_b
    -\left(k^2+\frac{1}{b^2}\right)\widehat Q&=0.
    \label{eq:F-Fourier-Bessel}
\end{align}
For $k>0$, regularity at $b=0$ and decay as $b\to\infty$ therefore select
the modes
\begin{equation}
    \begin{array}{lll}
    0<b<1:&
    \widehat P=S(k)I_0(kb),&
    \widehat Q=iS(k)I_1(kb),\\[2mm]
    b>1:&
    \widehat P=C(k)K_0(kb),&
    \widehat Q=-iC(k)K_1(kb).
    \end{array}
    \label{eq:F-Fourier-solutions}
\end{equation}
The boundary data of the instanton fix the amplitudes $S(k)$ and $C(k)$
given below. Substitution of Eq.~\eqref{eq:F-Fourier-solutions} into
Eq.~\eqref{eq:A1-before-parseval}, followed by Parseval's identity, gives
\begin{equation}
    A_1
    =
    \frac{1}{\pi^2}
    \int_0^\infty dk\,
    \left[
        S(k)^2\mathcal B_I(k)
        +C(k)^2\mathcal B_K(k)
    \right],
    \label{eq:A1-Bessel-representation}
\end{equation}
where
\begin{align}
    S(k)
    &=
    \frac{2(1+k)e^{-k}}{k}, \qquad
    C(k)=
    \frac{4}{\pi k}(k\cosh k-\sinh k),
    \\
    \mathcal B_I(k)
    &=
    \int_0^1db\left[
        1-I_0(kb)^2-I_1(kb)^2
        +\frac{2(1-b)}{b}I_1(kb)^2
    \right],
    \\
    \mathcal B_K(k)
    &=
    \int_1^\infty db\left[
        -K_0(kb)^2-K_1(kb)^2
        +\frac{2(1-b)}{b}K_1(kb)^2
    \right].
\end{align}
The Bessel identities
\begin{equation}
    \frac{d}{dx}(I_0^2-I_1^2)
    =
    \frac{2I_1^2}{x},
    \qquad
    \frac{d}{dx}(K_0^2-K_1^2)
    =
    \frac{2K_1^2}{x},
\end{equation}
give
\begin{equation}
    \mathcal B_I(k)
    =
    -\frac{2}{k}\int_0^k I_1(q)^2\,dq,
    \qquad
    \mathcal B_K(k)
    =
    -\frac{2}{k}\int_k^\infty K_1(q)^2\,dq.
\end{equation}
Moreover,
\begin{equation}
    S(k)
    =
    \sqrt{\frac{8k}{\pi}}K_{3/2}(k),
    \qquad
    C(k)
    =
    \sqrt{\frac{8k}{\pi}}I_{3/2}(k).
\end{equation}
Changing the order of integration in
Eq.~\eqref{eq:A1-Bessel-representation} therefore yields
\begin{equation}
    \pi^2A_1
    =
    -\frac{16}{\pi}\mathcal J,
\end{equation}
where
\begin{equation}
    \mathcal J
    =
    \int_{0<x<y}dx\,dy\,
    \left[
        I_1(x)^2K_{3/2}(y)^2
        +I_{3/2}(x)^2K_1(y)^2
    \right].
\end{equation}

To evaluate the remaining positive integral, set \(x=ty\), expand
\(I_\nu(ty)^2\), and use
\begin{equation}
    \int_0^\infty y^{s-1}K_\nu(y)^2\,dy
    =
    2^{s-3}
    \frac{
        \Gamma(s/2)^2
        \Gamma(s/2+\nu)
        \Gamma(s/2-\nu)
    }{\Gamma(s)}.
\end{equation}
The two contributions combine to
\begin{equation}
    \mathcal J
    =
    \pi\sum_{m=0}^\infty
    \left[
        \frac{(1/2)_m}{m!}
    \right]^2
    R_m,
\end{equation}
with
\begin{equation}
    R_m
    =
    \frac{
        (2m+1)(2m+5)(4m+7)(4m^2+14m+15)
    }{
        128(m+1)(m+2)^3(m+3)(2m+3)
    }.
\end{equation}
Partial fractions, together with the Euler beta representation of
\((1/2)_m/m!\), give
\begin{equation}
    \sum_{m=0}^\infty
    \left[
        \frac{(1/2)_m}{m!}
    \right]^2
    R_m
    =
    \frac{2\log2-1}{4}.
\end{equation}
Consequently,
\begin{equation}
    \mathcal J
    =
    \frac{\pi}{4}(2\log2-1),
    \qquad
    A_1
    =
    \frac{4(1-\log 4)}{\pi^2}.
\end{equation}

%%%%%%%%%%%%%
\section{Conjecture expansion at weak-coupling}\label{app:expansion}
To determine the first non-trivial correction in the small-$c$ expansion, we treat it by matched asymptotics and Wiener--Hopf factorization \cite{santana2026weakcouplinglimitlatticenonlinear, Tracy2016}. Using the main-text notation $r=c/q$, define
\begin{equation}
    K_r(x)=\log\left(1+\frac{r^2}{x^2}\right).
\end{equation}
We use the Fourier-transform convention
\begin{equation}
    \widehat f(k)=\int_{-\infty}^{\infty}e^{ikx}f(x)\,dx.
\end{equation}
Since
\begin{equation}
    \partial_r K_r(x)=\frac{2r}{x^2+r^2},
\end{equation}
and $K_0=0$, the Fourier transform of the logarithmic kernel is
\begin{equation}
    \widehat K_r(k)=\frac{2\pi}{|k|}\left(1-e^{-r |k|}\right).
\end{equation}
For $r|k|\ll1$, its reciprocal admits the expansion
\begin{equation}
    \frac{|k|}{2\pi(1-e^{-r |k|})} = \frac{1}{2\pi r}+\frac{|k|}{4\pi}+\cdots.
\end{equation}
To apply this expansion on the finite interval, we extend $g$ by zero outside $[-1,1]$. In real space, the operator with Fourier symbol $|k|$ is
\begin{equation}
    |D|h(t)=\frac{1}{\pi}\,\mathrm{P.V.}
    \int_{-\infty}^{\infty}
    \frac{h(t)-h(s)}{(t-s)^2}\,ds.
\end{equation}
Writing $g=r^{-1}g_{-1}+g_0+\cdots$, the first two orders of the integral equation give
\begin{equation}
    g_{-1}(t)=\frac{2t}{\pi},
    \qquad
    g_0(t)=\frac{1}{2}|D|g_{-1}(t).
\end{equation}
Evaluating the principal-value integral, we obtain the bulk expansion
\begin{equation}
    g(t) = \frac{2t}{\pi r}
    +\frac{1}{\pi^2}\log\frac{1+t}{1-t}
    +\frac{2t}{\pi^2(1-t^2)}+\cdots,
    \qquad 1-|t|\gg r.
\end{equation}
This expansion is nonuniform near $t=\pm1$, and the endpoint regions must therefore be treated separately. We introduce an intermediate cutoff $\varepsilon$ satisfying
\begin{equation}
    r\ll\varepsilon\ll\sqrt{r}.
\end{equation}
Using the oddness of $g$, the contribution from the bulk region is
\begin{equation}
    2\int_0^{1-\varepsilon}t g(t)\,dt = \frac{4}{3\pi r}-\frac{4\varepsilon}{\pi r}+\frac{2}{\pi^2}\log\frac{1}{\varepsilon}+\frac{2}{\pi^2}(\log2-1)+\cdots.
\end{equation}

We next resolve the boundary layer near $t=1$. Introduce the stretched coordinate $t=1-r\theta$ and define $H(\theta)=r g(1-r\theta)$. To leading order in $r$, the upper endpoint is mapped to a half-line and the integral equation becomes
\begin{equation}
    \int_0^\infty \log\left(1+\frac{1}{(\theta-\theta')^2}\right)H(\theta')d\theta' = 4.
\end{equation}
For the unit-width kernel, the full-line Fourier transform is $2\pi G(k)$, where
\begin{equation}
    G(k)=\frac{1-e^{-|k|}}{|k|}.
\end{equation}
We factorize $G$ as $G(k)=G_+(k)G_-(k)$, where $G_+$ and $G_-$ are analytic and nonzero in the upper and lower half-planes, respectively. Before fixing the exponential normalization, a convenient pair of factors is
\begin{equation}
    G_\pm^{(0)}(z)=\frac{1}{\Gamma(1\mp iz/2\pi)}
    \exp\left[\mp\frac{iz}{2\pi}\log(\mp iz)\right].
\end{equation}
The logarithms are analytic in the corresponding half-planes. For real $k$, the identity
\begin{equation}
    \frac{1}{\Gamma(1-ik/2\pi)\Gamma(1+ik/2\pi)}
    =\frac{\sinh(|k|/2)}{|k|/2}
\end{equation}
together with the factor $e^{-|k|/2}$ from the two exponentials shows that $G_+^{(0)}(k)G_-^{(0)}(k)=G(k)$. The factorization remains unchanged under $G_+\to e^{iaz}G_+$ and $G_-\to e^{-iaz}G_-$. We fix this freedom by requiring purely algebraic large-$|z|$ behavior. Stirling's expansion gives
\begin{equation}
    G_+^{(0)}(z)\sim\exp\left[-\frac{iz}{2\pi}(1+\log2\pi)\right](-iz)^{-1/2}.
\end{equation}
Thus, defining
\begin{equation}
    a=\frac{1+\log(2\pi)}{2\pi},
    \qquad
    G_+(z)=e^{iaz}G_+^{(0)}(z),
    \qquad
    G_-(z)=e^{-iaz}G_-^{(0)}(z),
\end{equation}
we obtain normalized factors satisfying
\begin{equation}
    G_\pm(z)\sim(\mp iz)^{-1/2}.
\end{equation}
For functions supported on the positive half-line, we write
\begin{equation}
    \widehat H_+(k)=\int_0^\infty e^{ik\theta}H(\theta)\,d\theta,
    \qquad \operatorname{Im}k>0.
\end{equation}
The standard Wiener--Hopf decomposition of the constant right-hand side then gives
\begin{equation}
    \widehat H_+(k)=\frac{2i}{\pi kG_+(k)}.
\end{equation}
To extract the large-$\theta$ behavior, we expand the normalized factor at small $k$:
\begin{equation}
    \log G_+(k)
    =-\frac{ik}{2\pi}\log(-ik)
    +\frac{ik}{2\pi}\left[1+\log(2\pi)-\gamma_{\mathrm E}\right]
    +O(k^2),
\end{equation}
where $\gamma_{\mathrm E}$ is Euler's constant. Consequently,
\begin{equation}
    \widehat H_+(k)
    =\frac{2i}{\pi k}
    -\frac{1}{\pi^2}\log(-ik)
    +\frac{1+\log(2\pi)-\gamma_{\mathrm E}}{\pi^2}
    +o(1).
\end{equation}
Equivalently, the integrated boundary solution has the large-$L$ expansion
\begin{equation}
    \int_0^L H(\theta)d\theta = \frac{2L}{\pi}+\frac{1}{\pi^2}\log L+\frac{1+\log2\pi}{\pi^2}+\cdots.
\end{equation}
The Euler constant in the small-$k$ transform cancels the Euler constant generated by the Laplace transform of $1/\theta$, leaving the constant displayed above.

Setting $L=\varepsilon/r$, the endpoint contribution is
\begin{equation}
    2\int_{1-\varepsilon}^1t g(t)dt = 2\int_0^{\varepsilon/r}(1-r\theta)H(\theta)d\theta.
\end{equation}
The term proportional to $r\theta$ contributes only $O(\varepsilon^2/r)+O(\varepsilon)$ and therefore vanishes for $r\ll\varepsilon\ll\sqrt{r}$. Keeping the terms that survive in this matching limit gives
\begin{equation}
    2\int_{1-\varepsilon}^1t g(t)dt = \frac{4\varepsilon}{\pi r}+\frac{2}{\pi^2}\log\frac{\varepsilon}{r}+\frac{2}{\pi^2}(1+\log2\pi)+o(1).
\end{equation}

Adding the bulk and boundary contributions, the dependence on the arbitrary matching scale $\varepsilon$ cancels, and we obtain
\begin{equation}
    \int_{-1}^1t g(t)dt = \frac{4}{3\pi r}+\frac{2}{\pi^2}\log\frac{4\pi}{r}+\cdots.
\end{equation}
To express this result in terms of the physical interaction parameter $\gamma_0$, we use $r/\gamma_0=\rho_0/q$ and the weak-coupling expansion of the Lieb equation \eqref{eq:Lieb-equation},
\begin{equation}
    \frac{r}{\gamma_0}
    =\frac{1}{4r}
    +\frac{1}{4\pi}\left[\log\frac{16\pi}{r}-1\right]
    +o(1).
\end{equation}
Inverting this relation gives
\begin{equation}
    r
    =\frac{\sqrt{\gamma_0}}{2}
    +\frac{\gamma_0}{8\pi}
    \left[\log\frac{32\pi}{\sqrt{\gamma_0}}-1\right]
    +o(\gamma_0).
\end{equation}
The dimensionless EFP rate function is related to the moment above by
\begin{equation}
    f(\gamma_0)
    =\frac{1}{2}\left(\frac{q}{\rho_0}\right)^2
    \int_{-1}^1t g(t)\,dt
    =\frac{1}{2}\left(\frac{\gamma_0}{r}\right)^2
    \int_{-1}^1t g(t)\,dt.
\end{equation}
Substituting the small-$r$ expansions and re-expanding for $\gamma_0\ll1$, all logarithms cancel at order $\gamma_0$, and we obtain
\begin{equation}
    f(\gamma_0)=\frac{16}{3\pi}\sqrt{\gamma_0}+\frac{4(1-\log 4)}{\pi^2}\gamma_0+o(\gamma_0).
\end{equation}
Numerically, the coefficient of the first correction is
\begin{equation}
    \frac{4(1-\log 4)}{\pi^2}\approx -0.15655920\ldots.
\end{equation}
This agrees with the coefficient obtained independently from the hydrodynamic perturbation theory.

%%%%%%%%%%%%%%%%%%%%%%%    
\end{appendix}
%%%%%%%%%%%%%%%%%%%%%%%
%%%%%%%%%%%%%%%%%%%%%%%

%%%%%%%%%%%%%%%%%%%%%%%%%%%%%%%%%%
%%%%%%%%%%%%%%%%%%%%%%%%%%%%
\end{document}